\documentclass[nofootinbib,aps,showpacs,preprint]{revtex4}
\usepackage{amsmath,chay}
\usepackage{citesort}
\usepackage{graphicx}
\newcommand{\eu}{\epsilon_{\mathrm{UV}}}
\newcommand{\ei}{\epsilon_{\mathrm{IR}}}
\begin{document}
\preprint{ \vbox{ 
\hbox{MIT-CTP 3904}}} 

\title{Transverse-momentum-dependent parton distribution function in
  soft-collinear effective theory}

\baselineskip 3.0ex

\author{Junegone Chay}\email{chay@korea.ac.kr} 

\bigskip

\affiliation{Center for Theoretical Physics, Laboratory for Nuclear
  Science, Massachusetts Institute of Technology, Cambridge, MA 02139}
\affiliation{Department
of Physics, Korea University, Seoul 136-701, Korea}

\begin{abstract}

\baselineskip 3.0ex 
Transverse-momentum-dependent parton distribution functions 
are analyzed in semi-inclusive deep inelastic scattering
at low transverse momentum using soft-collinear effective theory. The
transverse-momentum-dependent parton distribution functions are
defined on the lightcone without distorting the lightcone path nor adding
additional soft Wilson lines. In this approach, the comparison between
the integrated and unintegrated parton distribution functions becomes
transparent. The procedure of computing radiative
corrections in dimensional regularization is explained in detail, and
the divergence, which is a product of infrared and ultraviolet
divergence, is cancelled. The renormalization group equation for the
transverse-momentum-dependent parton distribution functions is derived.
It depends only on the relevant physical quantities and exhibits a
nontrivial scaling behavior because the longitudinal
momentum fraction and the transverse momentum are coupled in the
renormalization group equation. 
\end{abstract}


\maketitle

\section{Introduction}
High-energy scattering cross sections are usually expressed in a
factorized form, in which the hard scattering kernel and the parton
distribution functions (and the fragmentation functions if
semi-inclusive or exclusive scattering is considered) are
convoluted. The factorization property reflects the idea of separating
long-distance and short-distance physics in high-energy
scattering. The hard scattering occurs independent of the details
about how energetic partons are produced or how final-state
particles hadronize, and vice versa. Theoretically the factorization
implies that each factorized part can be treated separately. The
hard scattering kernel can be computed using perturbative QCD. Parton
distribution functions describe the nonperturbative nature of partons
to have a certain fraction of longitudinal momentum inside an incoming
hadron, hence not computable in perturbation theory. But their
evolution with respect to the factorization scale $\mu$ can be
computed. The factorization property is the basis to study many high
energy processes.

In inclusive high-energy processes, only integrated parton 
distribution functions (i.e., parton distribution functions
integrated over the transverse momentum) are needed to be convoluted
with the hard scattering kernel. The integrated
parton distribution functions can be well defined in a gauge-invariant
way as the matrix elements of the relevant operators and their
evolutions are described by the renormalization group
equation. If we consider semi-inclusive or exclusive
processes, that is, if one or more final-state particles are tagged,
more detailed information on the scattering such as the transverse
momentum dependence or spin correlations can be probed, but it is more
complicated than inclusive processes. In this
case, the scattering cross section 
is convoluted with one more nonperturbative fragmentation function,
which contains the information on the hadronization effect. 
However, since the transverse-momentum-dependent (TMD) parton
distribution function (PDF) was proposed 
\cite{Collins:1981uk,Collins:1981uw}, its definition 
and, accordingly, its radiative corrections are known to
cause severe problems \cite{Collins:2003fm}. These problems are
related to the failure of showing the cancellation of divergences
between real gluon emission and virtual gluon exchange in certain
calculational schemes once the TMD PDF is defined. The issues can be
categorized into the definition of the TMD PDF, the choice of gauge
(axial or covariant), and the regularization schemes though they are
intertwined with each other. In this
paper, the aspect of the transverse momentum dependence in
semi-inclusive deep inelastic scattering (SIDIS) is considered
and the spin correlation is not treated. TMD PDF has attracted a lot
of attention since its effect can be detected in experiments.

The problem can be described as follows \cite{Collins:2003fm}: The TMD
(or unintegrated) quark distribution function $f(x,{\bf k}_{\perp})$
at one loop is written as
\begin{equation} \label{csf}
f^{(1)} (x,{\bf k}_{\perp}) = P_{1R} (x,{\bf k}_{\perp}) -\delta (1-x)
\delta ({\bf k}_{\perp}) \int dx^{\prime} d^2 {\bf k}_{\perp}^{\prime}
P_{1R} (x^{\prime}, {\bf k}_{\perp}^{\prime}),  
\end{equation}
where $x$ is the lightcone momentum fraction and ${\bf k}_{\perp}$ is
the transverse momentum. The quantity $P_{1R}$ is the real gluon
emission amplitude at order $\alpha_s$, schematically given by 
\begin{equation} \label{csr}
P_{1R} (x,{\bf k}_{\perp}) \sim \alpha_s \Bigl[ \frac{1}{1-x}
  \frac{1}{{\bf k}_{\perp}^2 +\Delta^2} +\mathrm{regular \ as}\
  x\rightarrow 1 \Bigr], 
\end{equation}
where $\Delta^2$ involves the infrared regulator such as small quark
and gluon masses, and the second term in Eq.~(\ref{csf}) corresponds
to the virtual gluon 
emission. It is argued that for any ${\bf k}_{\perp}$, there is an
endpoint singularity as $x\rightarrow 1$, and there is no cancellation
between the real gluon emission and the virtual gluon contribution. If
the calculation is performed in light-cone gauge, the divergence
arises from the $1/k_+$ singularity in the gluon propagator. 

There have been several approaches to solving this problem. First,
Collins and Soper \cite{Collins:1981uk} proposed a non-gauge-invariant
definition for the TMD PDF, and chose the axial gauge $n\cdot A=0$,
but with a non-lightlike gauge fixing vector $n^{\mu}$. As a result, a
rapidity parameter $\zeta =(p\cdot n)^2/n^2$ ($n^2 \ne 0$) was
introduced and an additional evolution equation of the PDF on this
parameter was developed requiring that physical 
quantities like scattering cross sections are independent of this
scale. But the computation off the lightcone is complicated, and
the factorization property and the relation to the integrated PDF are
not clear in this approach \cite{Ji:2004wu}. Collins and Hautmann
\cite{Collins:1999dz}  suggested to put Wilson lines on the lightcone 
and redefine the TMD PDF by adding additional soft Wilson lines to
remove the divergence. They used method of regions to determine which
kinematic regions yield divergences, and found subtraction terms to
remove the divergence. From the explicit one-loop computation, they
found additional soft Wilson lines slightly off the lightcone to be
inserted in the gauge invariant definition of the TMD PDF. Cherednikov
and Stefanis \cite{Cherednikov:2007tw} redefined the TMD PDF by inserting
a transverse gauge link which provides an additional soft
counterterm. This additional term has the effect of cancelling the
unwanted divergence related to the cusped contour.     
All the approaches try to define TMD or unintegrated PDF
in a gauge-invariant way, and put some additional 
Wilson lines off the lightcone (taking the limit to the lightcone, if
possible) in order to remove the problematic singularity mentioned above
with the extra Wilson lines.

Soft-collinear effective theory (SCET)
\cite{Bauer:2000ew,Bauer:2000yr,Bauer:2001yt} is the theoretical
framework which is appropriate for describing energetic
particles. SCET has been applied successfully in $B$ decays and in
other high-energy processes \cite{Bauer:2002nz} such as deep inelastic 
scattering
\cite{Manohar:2003vb,Chay:2005rz,Becher:2006mr,Chen:2006vd,Idilbi:2007ff}.
It can be  also applied to SIDIS to describe the
TMD PDF. The factorization property of SIDIS cross sections can be
proved in a transparent way since the decoupling of collinear and soft
degrees of freedom can be made manifest in the formulation of SCET.
The TMD PDF can be defined in a gauge-invariant way putting
all the Wilson lines on the lightcone without introducing extra Wilson
lines off the lightcone. The computational technique
developed in SCET offers an improved regularization method to show
that the problem of the divergence is solved. The divergence appearing
in Eqs.~(\ref{csf}) and (\ref{csr}) is the mixture of ultraviolet and
infrared divergences, which should not be present for physical
quantities. In order to show the cancellation of the divergence 
using dimensional regularization, care must be taken to
extract $\delta (1-x)$ in Eq.~(\ref{csf}). In this
paper, the cancellation of the mixed divergence is explicitly
presented. In Section~\ref{kinematics}, kinematics relevant to SIDIS
is briefly considered and the current operators in SCET for SIDIS are
introduced. In Section~\ref{tmdpdf}, the TMD PDF and TMD soft Wilson
lines are defined in terms of the matrix elements of the relevant
operators. And the factorization of the hadronic tensor as a
convolution of the fragmentation function and the TMD PDF is
established. The TMD PDF itself is a convolution of the matrix
elements for the collinear operator and the soft Wilson line.
And the relation between the TMD and 
integrated PDF is explained. In Section~\ref{radcor}, the 
radiative corrections for the TMD collinear operator and the TMD soft
Wilson lines are computed at one loop. In Sec.~\ref{renorm}, the
renormalization group equations for the TMD collinear operator, the
TMD soft Wilson line, hence for the TMD PDF are presented. In
Section~\ref{conc}, a conclusion is given. 

\section{Kinematics \label{kinematics}}
In order to discuss TMD PDF, a reference frame should be selected
first to define what transverse momentum is referred to. In SIDIS such
as $e +p \rightarrow e + h+X$, where $h$ is an energetic tagged
final-state 
hadron and $X$ denotes all the remaining final states, the reference
frame is chosen such that the spatial components of the momentum
transfer $q^{\mu}$ from the leptonic system 
and the proton momentum $P^{\mu}$ lie in the $z$ axis.
In this frame, the momenta  $q^{\mu}$ and $P^{\mu}$ can be written as  
\begin{eqnarray} 
q^{\mu} &=& (\overline{n} \cdot q, {\bf q}_{\perp}, n\cdot q ) = (Q,
{\bf 0}_T, -Q), \nonumber \\ 
 P^{\mu} &=& P_{\bar{n}}^{\mu} =(\overline{n} \cdot P, {\bf P}_{\perp},
n\cdot P ) \sim (\Lambda^2 /Q, {\bf 0}_T , Q),  
\end{eqnarray}
where $q^2 =-Q^2$, and $\Lambda$ is a typical hadronic scale. The
lightlike vectors $n^{\mu}$ and $\overline{n}^{\mu}$ satisfy $n^2 =
\overline{n}^2 =0$, and $n\cdot \overline{n} =2$. We 
can set $n^{\mu} = (1,0,0,1)$ and $\overline{n}^{\mu} =
(1,0,0,-1)$, and the proton moves in the $\overline{n}$ ($-z$)
direction with this 
choice.  Note that an individual parton inside a proton can have 
nonzero transverse momentum of order $\Lambda$ due to the fluctuation
inside a proton, though the proton has no transverse momentum. We will
consider the case in which the transverse momentum of the final-state
particle is of order $\Lambda$. Of course, the outgoing hadron
can have transverse momentum much larger than $\Lambda$. But in this
case, we know from the momentum conservation that there should be other
final-state particles with large transverse momenta such that the 
final transverse momenta add up to be of order $\Lambda$. Technically,
as will be seen later, if the final hadron has a large transverse
momentum, it is extracted as a label momentum. Then the large
transverse momentum and the transverse momentum of order $\Lambda$ are
treated separately. Due to momentum conservation in each subspace
of the transverse momenta, the final result can be derived in a similar
way as in the case in which the final-state hadron is in the $n$
direction with small transverse momentum of order $\Lambda$. For this
reason, we consider the final-state hadron with $n$-collinear momentum
$P_n$ and small transverse momentum from now on for simplicity, noting
that the case with large transverse momentum can be treated in a
straightforward way.

At the hadronic level, we can define the invariants in
terms of $Q$, $P_{\bar{n}}$ and the outgoing hadron momentum
$P_n$ as
\begin{equation}
x_H = \frac{Q^2}{2P_{\bar{n}} \cdot q} = \frac{Q}{n\cdot P_{\bar{n}}},
\ \ z_H = \frac{P_{\bar{n}} \cdot P_n}{P_{\bar{n}} \cdot q} =
  \frac{\overline{n} \cdot P_n}{Q}. 
\end{equation}
Defining the momentum fractions $\xi$ and $\xi^{\prime}$ of the
partons to hadrons as $n\cdot p_{\bar{n}} = \xi n\cdot P_{\bar{n}}$
and $\xi^{\prime} \overline{n} \cdot p_n = \overline{n}\cdot P_n$, the
corresponding partonic invariants are given by
\begin{equation}
x = \frac{x_H}{\xi} = \frac{Q}{n\cdot p_{\bar{n}}}, \ \
  z=\frac{z_H}{\xi^{\prime}} = \frac{\overline{n} \cdot p_n}{Q}, 
\end{equation}
where the momenta of the incoming and outgoing partons are in the
$\overline{n}^{\mu}$, $n^{\mu}$ directions respectively such that
\begin{eqnarray}
p_{\bar{n}}^{\mu} &=& (\overline{n}\cdot p_{\bar{n}}, {\bf
  p}_{\bar{n}\perp}, n\cdot p_{\bar{n}}) \sim \Bigl( \displaystyle 
  \frac{x\Lambda^2}{Q}, \Lambda, \frac{Q}{x} \Bigr), \nonumber \\
p_n^{\mu} &=& (\overline{n} \cdot p_n , {\bf p}_{n\perp}, n\cdot p_n )
  \sim \Bigl( \displaystyle zQ, \Lambda, \frac{\Lambda^2}{zQ} \Bigr).
\end{eqnarray}
The scale $Q$ sets the large scale in the system, and there is a
hierarchy of scales in the momentum components such that
\begin{equation}
p_{\bar{n}}^{\mu} \sim Q(\lambda^2 , \lambda, 1), \ \ p_n^{\mu} \sim
Q(1,\lambda, \lambda^2),
\end{equation}
where $\lambda \sim \Lambda/Q$, and SCET describes the dynamics of
these particles. With this choice of the reference frame, the incoming
partons and the outgoing parton can have transverse momenta of order
$\Lambda$ with respect to the axis on which the incoming proton and
the photon lie. 

The electroproduction in SIDIS is mediated by the electromagnetic
current operator of the form $\overline{\psi}\gamma^{\mu} \psi$. The
corresponding electromagnetic current operator in SCET is given by
\begin{eqnarray} \label{emcur}
j_{\mu} (x) &=& C(Q,\mu) e^{i(\overline{n} \cdot p_n n\cdot x/2 -n\cdot
  p_{\bar{n}} \overline{n} \cdot x/2)} \overline{\chi}_n Y_n^{\dagger}
  \gamma_{\perp\mu} Y_{\bar{n}} \chi_{\bar{n}} (x), \nonumber \\
j^{\dagger}_{\mu} (x) &=&  C(Q,\mu) e^{-i(\overline{n} \cdot p_n n\cdot
  x/2 -n\cdot   p_{\bar{n}} \overline{n} \cdot x/2)}
  \overline{\chi}_{\bar{n}} \tilde{Y}^{\dagger}_{\bar{n}}
  \gamma_{\perp\mu} Y_n \chi_n (x), 
\end{eqnarray}
where $C(Q,\mu)$ is the Wilson coefficient obtained by matching the
current operator in the full theory onto SCET 
\cite{Manohar:2003vb} 
\begin{equation}
C(Q,\mu) = 1 +\frac{\alpha_s C_F}{4\pi} \Bigl( -\ln^2
\frac{Q^2}{\mu^2} +3 \ln \frac{Q^2}{\mu^2} -8 +\frac{\pi^2}{6}
\Bigr). 
\end{equation}
A collinear fermion field in the $n$ direction can be obtained from the
full theory as
\begin{equation}
\xi_n(x) = \sum_{\tilde{p}} e^{-i\tilde{p} \cdot x} \frac{\FMslash{n}
  \FMslash{\overline{n}}}{4} \psi (x), 
\end{equation}
where $\tilde{p}^{\mu} = \overline{n}\cdot p n^{\mu}/2$ is the large
label momentum. The field $\chi_n$, which is defined as $\chi_n =
W_n^{\dagger} \xi_n$, is introduced to simplify notation, and $W_n$ is
the collinear Wilson line given by 
\begin{equation}
W_n = \sum_{\mathrm{perm}} \exp \Bigl[ -g \frac{1}{\overline{n}\cdot
    \mathcal{P}} n\cdot A_n \Bigr],
\end{equation}
where $\overline{n}\cdot\mathcal{P}$ is the operator extracting the
label momentum. For $\overline{n}$ collinear particles, the
corresponding quantities are obtained by switching $n$ and
$\overline{n}$.

\begin{figure}[t] 
\begin{center}
\includegraphics[height=4.0cm]{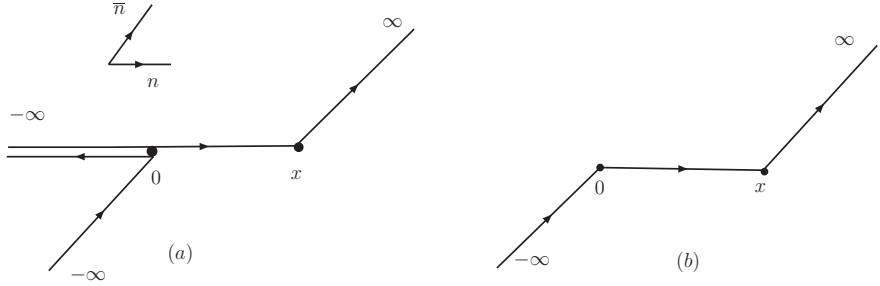}
\end{center}
\vspace{-0.5cm}
\caption{\baselineskip 3.0ex
The prescription for the soft Wilson lines in deep inelastic
scattering. (a) In the current $j_{\mu}^{\dagger}(x)$, $\xi_n$ from
$-\infty$ to $x$, and $\xi_{\bar{n}}$ from $x$ to $\infty$. In
$j_{\nu} (0)$, $\xi_{\bar{n}}$ from $-\infty$ to 0, and $\xi_n$ from 0
to $-\infty$. See Eq.~(\ref{wmunu}). (b) The net soft Wilson line
after the cancellation. \label{swilson} }    
\end{figure}

The collinear fields can be redefined to decouple from the soft fields
as \cite{Bauer:2001yt}
\begin{eqnarray}
&&\xi_n (x) \rightarrow Y_n \xi_n, \ A_n^{\mu} \rightarrow Y_n A_n^{\mu}
Y_n^{\dagger}  \nonumber \\
&&\xi_{\bar{n}} (x) \rightarrow Y_{\bar{n}} \xi_{\bar{n}}, \
A_{\bar{n}}^{\mu} \rightarrow Y_{\bar{n}}  A_{\bar{n}}^{\mu}
Y_{\bar{n}}^{\dagger},
\end{eqnarray}
where $Y_n$ and $Y_{\bar{n}}$ are soft Wilson lines. The form and the
analytic structure of the soft Wilson lines can be determined from the
direction of the underlying collinear particles, and depend on whether
the collinear particle is a particle or an antiparticle
\cite{Chay:2004zn}. And the basic building blocks for the soft Wilson
lines are given as
\begin{equation}
Y_n (x)= P\exp \Bigl[ ig \int_{-\infty}^x ds n\cdot A_s (ns)\Bigr], \
\tilde{Y}_n (x) = \overline{P} \exp \Bigl[ -ig \int_x^{\infty} ds
  n\cdot A_s (ns)\Bigr], 
\end{equation}
and their hermitian conjugates, where $P$ and $\overline{P}$ denote
path and anti-path ordering respectively. In Eq.~(\ref{emcur}), 
the soft Wilson lines are chosen appropriate for deep inelastic
scattering, as explained in detail in Ref.~\cite{Chay:2004zn}. The
prescription of the soft Wilson lines is shown in Fig.~\ref{swilson}
(a), which is equal to Fig.~\ref{swilson} (b) since the overlapping
Wilson lines are cancelled. 

\begin{figure}[b] 
\begin{center}
\includegraphics[height=4.0cm]{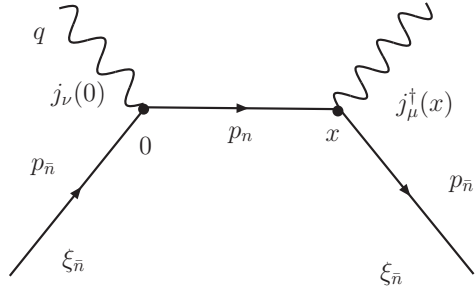}
\end{center}
\vspace{-0.5cm}
\caption{\baselineskip 3.0ex
Feynman diagram for forward Compton scattering amplitude in deep
inelastic scattering. The scattering cross section is proportional to
the discontinuity of the intermediate states by the optical
theorem.\label{dis} }    
\end{figure}

There exist other equivalent prescriptions for the soft Wilson
lines, but the procedure of calculating anomalous dimensions should
also be prescribed. The above prescription is for the forward Compton
scattering amplitude in deep inelastic scattering. (See
Fig.~\ref{dis}.) This prescription is simple in the sense that 
the same Feynman rules can be used anywhere in  Feynman diagrams,
and it makes the calculation of the anomalous dimension
straightforward. Note that the choice of the directions for the
underlying collinear particles such as forward scattering amplitude
applies only to the prescription for the soft Wilson lines, 
not to the actual processes.  In other prescriptions, the same result
can be reached, but by a different procedure. For example, if soft
Wilson lines are prescribed with a matrix element squared for a scattering
cross section instead of the time-ordered product for the forward
scattering amplitude, different Feynman rules should be applied across
the cut to produce the desired result \cite{Collins:1981uk}, or the
cusp angle in computing the anomalous dimension should be inverted in
the complex conjugate part at the final step \cite{Korchemsky:1992xv}.

\section{TMD parton distribution functions and soft Wilson
  lines\label{tmdpdf}} 
Consider the hadronic tensor, which is defined as
\begin{equation} \label{wmunu}
W_{\mu\nu} = \sum_X \int d^4 x e^{iq\cdot x} \langle
P | j_{\mu}^{\dagger} (x) | X h\rangle \langle Xh | j_{\nu}
(0)|P\rangle, 
\end{equation}
where the sum is over all the final states $X$ except
the tagged hadron $h$. The SIDIS cross section is proportional to the
hadronic tensor $W_{\mu\nu}$. It can be written in terms of the
SCET current operators. The exponential factors extracting the label
momenta all 
cancel due to momentum conservation requiring $x=z=1$ (or $\xi = x_H$,
$\xi^{\prime} = z_H$), and it becomes
\begin{eqnarray} \label{wmunu1}
W_{\mu\nu} &=& C^2 (Q) \int d^4 x \sum_X \langle P_{\bar{n}} |
\overline{\chi}_{\bar{n}} \tilde{Y}^{\dagger}_{\bar{n}}
\gamma_{\perp\mu} Y_n \chi_n (x) |X h\rangle \langle Xh|\overline{\chi}_n
Y_n^{\dagger} \gamma_{\perp\nu} Y_{\bar{n}} \chi_{\bar{n}} (0)
|P_{\bar{n}}\rangle \nonumber \\
&=&  C^2 (Q) \int d^4 x \sum_X \langle P_{\bar{n}} |
\Bigl(\overline{\chi}_{\bar{n}}\gamma_{\perp\mu} \frac{\FMslash{n}}{2}
\Bigr)_i^a \Bigl(\tilde{Y}^{\dagger}_{\bar{n}}
Y_n\Bigr)^{ab} \Bigl( \frac{\FMslash{\overline{n}}}{2} \chi_n
\Bigr)_i^b  (x) |X h\rangle \langle Xh|\Bigl(\overline{\chi}_n
\Bigr)_k^c \nonumber \\
&&\times \Bigl(Y_n^{\dagger} Y_{\bar{n}} \Bigr)^{cd} \Bigl(
\gamma_{\perp\nu}\chi_{\bar{n}} \Bigr)_k^d (0) |P_{\bar{n}}\rangle   
\end{eqnarray}
where the color indices ($a$, $b$, $c$, $d$) and the Dirac indices
($i$, $j$, $k$) are explicitly shown in the last expression. The state 
$|P_{\bar{n}}\rangle$ denotes the proton state in SCET, which consists
of $\overline{n}$-collinear partons.
 
Since there are no $\overline{n}$-collinear particles in the final
state, the fragmentation function $D(z_H, n\cdot r,  r_{\perp})$ is
defined as the average of the following matrix element over color and
Dirac indices such that 
\begin{equation} \label{tmdfrag}
\frac{1}{2N}\mathrm{Tr} \sum_X \langle 0|\Bigl(
\frac{\FMslash{\overline{n}}}{2} \chi_n (x) \Bigr)_i^b |Xh\rangle
\langle Xh| \Bigl(\overline{\chi}_n (0) \Bigr)^c_k |0\rangle
 = \delta^{bc} \delta_{ik} \int \frac{d^4 r}{(2\pi)^4} e^{-i r\cdot x}
 D (z_H,n\cdot r,  r_{\perp}),   
\end{equation}
where $D(z_H,n\cdot r, r_{\perp})$ depends only on $n\cdot r$ and 
$r_{\perp}$ and the label momentum is given by $\overline{n} \cdot p_n
= \overline{n} \cdot P_n /z_H$. The definition of the fragmentation
function also holds when the final-state hadron $h$ has large
transverse momentum after 
summing over all the final-state particles, though the partonic
variables $x$ and $z$ can have values less than 1, and appropriate
adjustments should be made accordingly. Note that, as in inclusive
deep inelastic scattering, when the contribution of 
the final-state particles is all summed over, there is no dependence
on the transverse momentum and it becomes the jet function $J_P
(n\cdot r)$ \cite{Chay:2005rz}, which depends only on $n\cdot
r$. Therefore the transverse-momentum 
dependence in the fragmentation function, and subsequently in the
PDF appears only when we specify at least one final-state particle
with nonzero transverse momentum. From now on, the subscript in
$z_H$ (and $x_H$) is dropped for simplicity with the
understanding that the momentum fractions refer to the hadronic
variables.

Using the fragmentation function, the hadronic tensor
$W_{\mu\nu}=-g^{\perp}_{\mu\nu}W$ can be written, by integrating 
over $\overline{n} \cdot r$, as
\begin{eqnarray} \label{wscet}
W&=& \int d\omega C^2 (\omega) \int \frac{d\overline{n}\cdot x d^2
  x_{\perp}}{2 (2\pi)^3} \int d n\cdot r d^2 r_{\perp} e^{-i(n\cdot r
  \overline{n}\cdot x/2 + r_{\perp} \cdot x_{\perp})} D (z,n\cdot r, 
  r_{\perp}) \nonumber \\
&\times& \langle P_{\bar{n}}| \overline{\chi}_{\bar{n}} (\overline{n}
  \cdot x, x_{\perp}) 
 \delta (\omega -\mathcal{P}_+)  \frac{\FMslash{n}}{2} \chi_{\bar{n}}
  (0)|P_{\bar{n}} \rangle 
  \frac{1}{N} \langle 0| \mathrm{tr} \, \tilde{Y}^{\dagger}_{\bar{n}}
  Y_n   (\overline{n}\cdot x, x_{\perp}) Y_n^{\dagger} Y_{\bar{n}} (0)
  |0\rangle, 
\end{eqnarray}
where the soft Wilson lines are extracted from the $\overline{n}$
states, since they are decoupled from the collinear part, and the
color projection is taken. In Eq.~(\ref{wscet}), the delta function
$\delta (\omega -\mathcal{P}_+)$ is inserted explicitly with
$\mathcal{P}_+ = n\cdot \mathcal{P} +n\cdot \mathcal{P}^{\dagger}$ 
since the Wilson coefficient is actually an operator $C(n\cdot
\mathcal{P}, n\cdot \mathcal{P}^{\dagger})$. By expanding the
operators with respect to $x_{\perp}$ and $\overline{n}\cdot x$, we
obtain 
\begin{eqnarray}
\overline{\chi}_{\bar{n}} (0, x_{\perp})\frac{\FMslash{n}}{2}  
 \delta (\omega -\mathcal{P}_+)   \chi_{\bar{n}}
  (0) &=& \int d^2 k_{\perp}  e^{-i k_{\perp} \cdot x_{\perp}}
 \overline{\chi}_{\bar{n}}  \frac{\FMslash{n}}{2}  
 \delta (\omega -\mathcal{P}_+)  \delta^{(2)} (k_{\perp}
 +i\partial_{\perp})  \chi_{\bar{n}}(0), \nonumber \\
\frac{1}{N}  \mathrm{tr} \, 
\tilde{Y}^{\dagger}_{\bar{n}} Y_n (\overline{n}\cdot x, x_{\perp})
Y_n^{\dagger} Y_{\bar{n}} (0) 
&=& \int d\eta d^2 s_{\perp} e^{-i(\eta \overline{n}\cdot x/2
  +s_{\perp} \cdot x_{\perp})} \nonumber \\
&\times& \frac{1}{N}
\mathrm{tr} \, \tilde{Y}^{\dagger}_{\bar{n}} Y_n \delta (\eta +n\cdot
i\partial  )
\delta^{(2)} (s_{\perp} + i\partial_{\perp})  Y_n^{\dagger}
Y_{\bar{n}} (0). 
\end{eqnarray}
Note that the dependence of $\overline{\chi}_n$ on $\overline{n} \cdot
x$ is dropped since the label momentum is already extracted, and the
remaining momentum can be neglected. Then $W$ is written as
\begin{eqnarray} \label{finalw}
W&=& \int d\omega C^2 (\omega) \int \frac{d\overline{n}\cdot x d^2
  x_{\perp}}{2 (2\pi)^3} \int d n\cdot r d^2 r_{\perp}  \int d^2
  k_{\perp} \int d\eta d^2 s_{\perp} \nonumber \\
&\times& e^{-i(n\cdot r
  \overline{n}\cdot x/2 + r_{\perp} \cdot x_{\perp})} e^{-ik_{\perp}
  \cdot x_{\perp}} e^{-i(\eta \overline{n}\cdot x/2 +s_{\perp} \cdot
  x_{\perp})}  
  D (z,n\cdot r,  r_{\perp})  \langle
  P_{\bar{n}}| \mathcal{O}_c   (\omega, k_{\perp}) |P_{\bar{n}}\rangle
  K(\eta, s_{\perp}) \nonumber \\
&=& \int d\omega C^2 (\omega) \int d n\cdot r d^2 r_{\perp}  \int d^2
  k_{\perp} \int d\eta d^2 s_{\perp} \delta (\eta +n\cdot r)
  \delta^{(2)} (r_{\perp} +k_{\perp}+ s_{\perp}) \nonumber \\
&\times& D (z,n\cdot r,  r_{\perp})  \langle
  P_{\bar{n}}| \mathcal{O}_c   (\omega, k_{\perp}) |P_{\bar{n}}\rangle
  K(\eta, s_{\perp}),
\end{eqnarray}
where the TMD collinear operator $\mathcal{O}_c (\omega, k_{\perp})$
and the TMD soft Wilson line $S(\eta, s_{\perp})$ are defined as
\begin{eqnarray} \label{tmdop}
\mathcal{O}_c (\omega, k_{\perp}) &=& \overline{\chi}_{\bar{n}}
\frac{\FMslash{n}}{2} \delta (\omega -\mathcal{P}_+ ) \delta^{(2)}
(k_{\perp} + i\partial_{\perp}) \chi_{\bar{n}}, \nonumber \\
S(\eta,  s_{\perp}) &=& \frac{1}{N} \mathrm{tr} \,
\tilde{Y}_{\bar{n}}^{\dagger} 
Y_n \delta (\eta +n\cdot i\partial) \delta^{(2)} (
  s_{\perp}+i\partial_{\perp}) Y_n^{\dagger} Y_{\bar{n}},  
\end{eqnarray}
and $K(\eta,s_{\perp}) =\langle 0| S(\eta,s_{\perp})|0\rangle$. 

Note that the collinear particles and the soft Wilson
line are not on the light cone $x^{\mu} = n^{\mu} (\overline{n}\cdot
x)/2$ in Eq.~(\ref{wscet}), but are slightly off the light cone in the
transverse direction $x_{\perp}$. It is the expression with which
all the approaches on TMD PDF agree, but the treatment of the TMD PDF
and soft Wilson lines takes  different paths from this
point. Previously, these quantities were manipulated at this stage 
since it would  presumably cause a serious problem by putting   
the particles on the lightcone. Here all the particles are on the
light cone, as in  Eq.~(\ref{finalw}), and this is the starting point
in this approach. It will be shown later that the radiative
computations can be performed after putting all the collinear
particles on the lightcone. That is, all the collinear particles are
on the lightcone, so are the collinear and soft Wilson lines by
expanding about the lightcone. As a result, Eq.~(\ref{tmdop}) is used for
the  TMD collinear operator, and the TMD soft Wilson line.

The hadronic tensor can be expressed in terms of the TMD PDF.
To define TMD PDF in SCET, let us recall the definition in the full
theory, which is given as  
\begin{equation} \label{fullpdf}
f(y,p_{\perp} ) = \int \frac{d\overline{n} \cdot x d^2 x_{\perp}}{2
  (2\pi)^3} e^{-i(y n\cdot P \overline{n} \cdot x/2 +p_{\perp} \cdot
  x_{\perp})} \langle P | \overline{\psi} (\overline{n} \cdot x,
  x_{\perp}) W[(\overline{n}\cdot x, x_{\perp}),0]
  \frac{\FMslash{n}}{2} \psi (0) |P\rangle, 
\end{equation}
where $W[(\overline{n}\cdot x, x_{\perp}),0]$ is the Wilson line
connecting the two points to make a gauge-invariant operator, and $y$
is the longitudinal momentum fraction of the parton $n\cdot
p_{\bar{n}} = y n\cdot P_{\bar{n}}$. It is straightforward to write
this definition in SCET by replacing the quark fields with the
collinear fields and the 
Wilson lines with the corresponding collinear and soft Wilson
lines. However, there is one more modification necessary, which
is due to the way SCET describes disparate scales. Physically, the
parton with the momentum fraction $y$ can emit soft gluons before
undergoing a hard collision, and the longitudinal momentum of the
parton in a hard collision is given by $yn\cdot 
P_{\bar{n}} -\kappa$, where $\kappa$ is the longitudinal momentum of
the soft gluons. In general, $\kappa$ is negligible since it is of
order $\Lambda$, but it should be kept to guarantee the momentum
conservation in each subspace of the momentum region. Note that      
the label momentum of order $Q$ and the residual fluctuation of order
$\Lambda$ in SCET reside in different subspaces, and that is
reflected, for example, in the treatment of the integral
\begin{equation} \label{deli}
\int dz e^{i (\omega - n\cdot p)z} e^{i(\kappa -\eta)z} =
\delta_{\omega, n\cdot p} \int dz e^{i(\kappa -\eta)z} =
\delta_{\omega, n\cdot p} (2\pi) \delta (\kappa -\eta), 
\end{equation}
in which the integration over the label momenta of order $Q$ ($\omega$
and $n\cdot p$) yields a Kronecker delta, but the integration over the
residual momenta of order $\Lambda$ ($\kappa$ and $\eta$) yields a
delta function. 

The TMD PDF in SCET is obtained from Eq.~(\ref{fullpdf}) by switching
to the fields in SCET, and by replacing $yn\cdot P_{\bar{n}}$ with
$yn\cdot P_{\bar{n}} -\kappa$.  As a result, the TMD PDF in SCET
depends on an additional scale $\kappa$, which describes soft gluon
emission, and it is defined as 
\begin{eqnarray} \label{tmdc1}
f(y,p_{\perp},\kappa ) &=& \int d\omega \int \frac{d\overline{n} \cdot x
  d^2   x_{\perp}}{2 
  (2\pi)^3} e^{i(\omega/2 - y n\cdot P_{\bar{n}}+\kappa) \overline{n}
  \cdot   x/2 -ip_{\perp} \cdot   x_{\perp}} \nonumber \\ 
&\times& \langle P_{\bar{n}} | \overline{\chi}_{\bar{n}}
  (x_{\perp}) \delta (\omega -\mathcal{P}_+) \frac{\FMslash{n}}{2}
  \chi_{\bar{n}} (0) |P_{\bar{n}}   \rangle   \frac{1}{N} \langle 0|
  \mathrm{tr} \, \tilde{Y}_{\bar{n}}^{\dagger} Y_n (\overline{n}\cdot
  x, x_{\perp}) Y_n^{\dagger} Y_{\bar{n}}(0) |0\rangle,
\end{eqnarray}
where  $\omega =2 n\cdot p_{\bar{n}}$ comes from the label momentum,
and the soft Wilson lines are decoupled from the collinear sector. 
The dependence on $\overline{n} \cdot x$ in $\chi_{\bar{n}}$ is
neglected because the remaining momentum of the collinear fermion in
the $\overline{n}$ direction is much smaller than the soft momentum in the
soft Wilson lines. By expanding with respect to $x_{\perp}$,
$f(y,p_{\perp},\kappa)$ is written as 
\begin{eqnarray} \label{tmdc2}
f(y,p_{\perp},\kappa) &=& \int d\omega \int \frac{d\overline{n}\cdot x
  d^2   x_{\perp}}{2 (2\pi)^3} e^{i (\omega/2 - y n\cdot
  P_{\bar{n}})\overline{n} \cdot x/2}
  e^{i(\kappa \overline{n}\cdot x/2 +p_{\perp}\cdot x_{\perp})}
  \nonumber \\
&\times& \int d^2 k_{\perp} e^{-i k_{\perp} \cdot x_{\perp}} 
\langle  P_{\bar{n}} | \mathcal{O}_c (\omega, k_{\perp}) 
|P_{\bar{n}}\rangle \int d\eta d^2 s_{\perp} 
  \, e^{-i \eta \overline{n}\cdot x/2 -is_{\perp} \cdot x_{\perp}} 
K(\eta,s_{\perp}) \nonumber \\
&=& \int d\omega \int d^2 k_{\perp} \int d\eta d^2 s_{\perp}
  \delta_{\omega, 2yn\cdot P_{\bar{n}}}   \delta
  (\kappa -\eta) \delta^{(2)} (p_{\perp} +k_{\perp} + s_{\perp})
  \nonumber \\
&\times&   \langle  P_{\bar{n}} | \mathcal{O}_c (\omega, k_{\perp}) 
|P_{\bar{n}}\rangle K(\eta,s_{\perp}), 
\end{eqnarray}
where Eq.~(\ref{deli}) is used. From the delta function $\delta
(\kappa-\eta)$, it is clearly seen that $\kappa$ describes the soft
gluon emission. To simplify further, let us define the matrix element
of  $\mathcal{O}_c (\omega, k_{\perp})$ as
\begin{eqnarray} \label{gdef}
&& \langle P_{\bar{n}} |
\overline{\chi}_{\bar{n}} \frac{\FMslash{n}}{2}  \delta (\omega
-\mathcal{P}_+) \delta^{(2)} ( k_{\perp} + i\partial_{\perp})
\chi_{\bar{n}} (0) |P_{\bar{n}}\rangle \nonumber \\
&& = \int du \int d^2 \rho_{\perp}
\delta \Bigl(u- 
\frac{\omega}{2n\cdot P_{\bar{n}}} \Bigr) \delta^{(2)} (k_{\perp}
+\rho_{\perp} ) g(u, \rho_{\perp}).   
\end{eqnarray}
Plugging this expression into Eq.~(\ref{tmdc2}) and integrating over
the delta functions, the TMD PDF is a convolution of the matrix
elements of the TMD collinear operator and soft Wilson lines, and is
written as 
\begin{equation} \label{fkappa}
f(y,p_{\perp},\kappa) = \int d^2 \rho_{\perp} d^2 s_{\perp}
\delta^{(2)} (p_{\perp} -\rho_{\perp}+s_{\perp} ) g(y,\rho_{\perp})
K(\kappa, s_{\perp}). 
\end{equation}
From the delta function, we can see that the transverse momentum of
the parton $p_{\perp}$ is given by the original transverse momentum
$\rho_{\perp}$ of the collinear parton before the soft gluon emission,
subtracted by the transverse momentum of soft gluons $s_{\perp}$
($p_{\perp} = \rho_{\perp} -s_{\perp}$).  
From Eq.~(\ref{gdef}), the relation between
$g(y,\rho_{\perp})$ and $\mathcal{O}_c (\omega, k_{\perp})$ is given
by
\begin{equation} \label{gk}
g(y,-k_{\perp}) =\langle P_{\bar{n}}| \mathcal{O}_c (\omega =2yn\cdot
P_{\bar{n}}, k_{\perp})|P_{\bar{n}}\rangle. 
\end{equation}
  
In terms of the TMD PDF, the hadronic quantity $W$ is written as
\begin{eqnarray} \label{wfin}
W(y,z) &=& \int d\omega C^2 (\omega) \int du \delta \Bigl( u -
\frac{\omega}{2n\cdot P_{\bar{n}}} \Bigr) \int dn\cdot r d^2 r_{\perp}
\int d\eta \delta (n\cdot r +\eta) D(z,n\cdot r, r_{\perp}) \nonumber
\\ 
&\times& \int d^2 s_{\perp} d^2 \rho_{\perp} \delta^{(2)} (r_{\perp} +
s_{\perp} -\rho_{\perp}) g(u,\rho_{\perp}) K(\eta, s_{\perp})
\nonumber \\
&=& C^2 (n\cdot p_{\bar{n}}) \int d\eta d^2 r_{\perp} D(z,-\eta,
r_{\perp}) f(u,r_{\perp},\eta), 
\end{eqnarray}
where Eq.~(\ref{fkappa}) and the fact that $\omega = 2n\cdot
p_{\bar{n}} = 2yn\cdot P_{\bar{n}}$ are used in the last expression. 
$W$ factorizes into the TMD fragmentation and the TMD PDF, and the TMD
PDF is factorized in terms of a convolution of the collinear and the
soft parts, as given in Eq.~(\ref{gdef}).
Eqs.~(\ref{fkappa}) and (\ref{wfin}) are the main result which proves
the factorization of the hadronic tensor. In
Eq.~(\ref{wfin}),  the energy-momentum conservation is manifest 
from the delta functions. The small momentum components in the
$\overline{n}$ direction in the hadron and in the soft gluons are
compensated by each other ($n\cdot r + \eta =0$), and the transverse
momentum $\rho_{\perp}$ from the incoming parton is equal to the sum
of those for the outgoing hadron ($r_{\perp}$) and the soft gluons
($s_{\perp}$) with  $\rho_{\perp}= r_{\perp} + s_{\perp}$.

It is useful to consider the relations between the TMD PDF and soft
Wilson lines and the integrated PDF and soft Wilson lines, which
appear in inclusive deep inelastic scattering. 
In SCET, the relation between the TMD PDF and the integrated PDF can
be established in a transparent way. The integration of
$f(y,p_{\perp},\kappa)$  
over the transverse momentum can be performed easily
when we consider the Fourier transform (or the impact parameter space 
representation) for $g(y,\rho_{\perp})$ and $K(\eta,s_{\perp})$ 
\begin{eqnarray} \label{impar}
\tilde{g} (y,b_{\perp}) &=& \int d^2 \rho_{\perp} e^{ib_{\perp} \cdot
  \rho_{\perp}} g(y,\rho_{\perp}), \ \ g(y,\rho_{\perp}) = \int
  \frac{d^2 b_{\perp}}{(2\pi)^2} e^{-ib_{\perp} \cdot \rho_{\perp}}
  \tilde{g} (y,b_{\perp}), \nonumber \\
\tilde{K} (\kappa, c_{\perp}) &=& \int d^2 s_{\perp}
  e^{ic_{\perp}\cdot s_{\perp}} K(\kappa,c_{\perp}), \ K(\kappa,
  s_{\perp}) = \int  \frac{d^2 c_{\perp}}{(2\pi)^2} e^{-ic_{\perp}
  \cdot s_{\perp}}   \tilde{K} (\kappa,c_{\perp}). 
\end{eqnarray}
Then the impact parameter space representation of the TMD PDF is
written as
\begin{equation} \label{fimp}
\tilde{f} (y,b_{\perp},\kappa)= \int d^2 p_{\perp} e^{i b_{\perp}
  \cdot p_{\perp}} f(y,p_{\perp},\kappa)
= \tilde{g} (y,b_{\perp}) \tilde{K}
(\kappa,-b_{\perp}).  
\end{equation}
By putting $b_{\perp}=0$ in Eq.~(\ref{fimp}), we have
\begin{equation}
f(y,\kappa) = \int d^2 p_{\perp} f(y,p_{\perp},\kappa) = \int d^2
\rho_{\perp} g(y,\rho_{\perp}) \int d^2 s_{\perp} K(\kappa,
s_{\perp}) =g(y)K(\kappa), 
\end{equation}
where the integrated collinear matrix element $g(y)$ and the soft
Wilson line $K(\eta)$ are given in SCET as \cite{Chay:2005rz} 
\begin{eqnarray}
g(y) &=& \int d\omega \delta_{\omega, 2y n\cdot P_{\bar{n}}} \langle
P_{\bar{n}} | \overline{\chi}_{\bar{n}} \frac{\FMslash{n}}{2} \delta
(\omega -\mathcal{P}_+) \chi_{\bar{n}}|P_{\bar{n}} \rangle =\int d^2
\rho_{\perp} g(y,\rho_{\perp}),  \nonumber \\ 
K(\eta) &=& \frac{1}{N} \langle 0| \mathrm{tr} \,
\tilde{Y}^{\dagger}_{\bar{n}} Y_n \delta (\eta +n\cdot i\partial  )  
Y_n^{\dagger} Y_{\bar{n}} (0) |0\rangle =\int d^2 s_{\perp}
K(\eta,s_{\perp}).
\end{eqnarray}
It means that the integrated PDF is obtained by the product of the
integrated collinear part and the integrated soft Wilson line to all
orders in $\alpha_s$. The explicit relation at one loop between TMD
and integrated PDF and soft Wilson lines will be presented in
Sec.~\ref{renorm}.

Since the relation between the TMD PDF and the TMD collinear and soft
Wilson line operators is established, we can focus on each operator
and compute its radiative corrections. Note that the soft Wilson lines
are present almost everywhere in high-energy processes though their
analytic structure differs depending on the physical
processes. Furthermore, as can be seen in Eq.~(\ref{wmunu1}), part of
the soft Wilson line ($Y_n$) comes from the collinear fields
contributing to the fragmentation functions. Therefore it is
hard to include the soft Wilson line residing only in the PDF. Instead
they are sometimes regarded as universal objects,   
and are pulled out of the collinear part. And the collinear part only is
referred to as contributing to the PDF. However, to make the
comparison with the integrated PDF easy, the soft Wilson line is
included in the PDF, which is consistent with 
the definition in the full theory. Therefore the distinction between
the PDF and the collinear matrix element should be carefully
understood from the context in the literature.

\section{Radiative corrections\label{radcor}}
The relevant  collinear  and soft Wilson line operators in considering
the TMD PDF are given as
\begin{eqnarray}
\mathcal{O}_c (\omega, {\bf k}_{\perp}) &=& \overline{\chi}_{\bar{n}}
\frac{\FMslash{n}}{2} \delta (\omega -\mathcal{P}_+ ) \delta^{(2)}
({\bf k}_{\perp} + i\nabla_{\perp}) \chi_{\bar{n}}, \nonumber \\
S(\eta, {\bf s}_{\perp}) &=& \frac{1}{N} \mathrm{tr} \,
\tilde{Y}_{\bar{n}}^{\dagger} 
Y_n \delta (\eta +n\cdot i\partial) \delta^{(2)} ({\bf
  s}_{\perp}+i\nabla_{\perp}) Y_n^{\dagger} Y_{\bar{n}}.  
\end{eqnarray}
From now on, the transverse momentum is written as a 3-vector. The
radiative corrections for these operators and their renormalization
group equations will be considered.

\subsection{One loop correction of the collinear operator}

\begin{figure}[b] 
\begin{center}
\includegraphics[height=6.7cm]{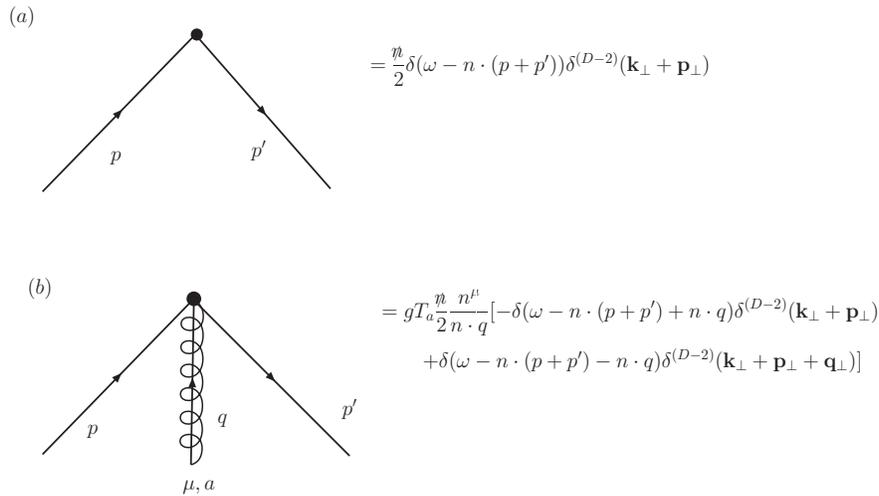}
\end{center}
\vspace{-0.5cm}
\caption{\baselineskip 3.0ex
Feynman rules for the collinear operator $\mathcal{O}_c (\omega,
{\bf k}_{\perp})$ to order $g$. In (b), $T_a$ are $SU(3)$ color
generators, and the momentum of the collinear gluon is incoming.
\label{cofeyn} }   
\end{figure}

\begin{figure}[t] 
\begin{center}
\includegraphics[width=11.0cm]{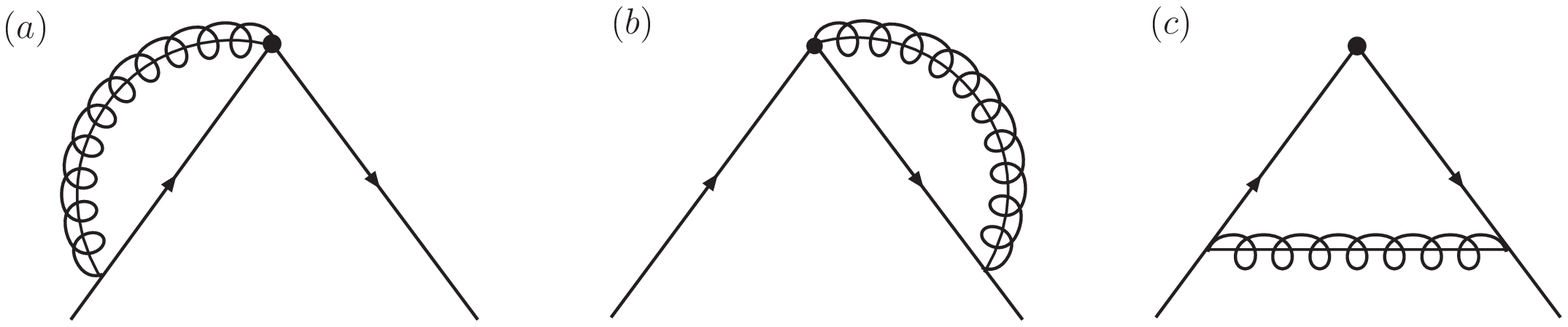}
\end{center}
\vspace{-0.5cm}
\caption{\baselineskip 3.0ex
Feynman diagrams of the radiative correction for the collinear
operator $\mathcal{O}_c$ at one loop. \label{corad} }   
\end{figure}

Let us consider the radiative corrections for the collinear operator
\begin{equation} \label{opoc}
\mathcal{O}_c (\omega, {\bf k}_{\perp}) = \overline{\chi}_{\bar{n}}
\frac{\FMslash{n}}{2} \delta (\omega -\mathcal{P}_+) \delta^{(D-2)}
({\bf   k}_{\perp} +i{\bf \nabla}_{\perp} ) \chi_n.
\end{equation}
The Feynman rules for the operator $\mathcal{O}_c$ to order
$g$ is shown in Fig.~\ref{cofeyn} by expanding the collinear Wilson
lines $W_{\bar{n}}$. The Feynman diagrams for the
radiative correction at one loop is shown in Fig.~\ref{corad}. The
computation is performed with dimensional regularization in
$D=4-2\epsilon$ dimensions for the ultraviolet divergence and the
nonzero external $p^2$ acts as an infrared cutoff. Note that the
two-dimensional delta function in Eq.~(\ref{opoc}) becomes
$(D-2)$-dimensional delta function in dimensional
regularization. We can also use the two-dimensional delta
function. While the above definition satisfies the rotational
invariance in the $(D-2)$ transverse plane, the two-dimensional
delta function breaks this symmetry. Though the symmetry is broken
using the two-dimensional delta functions, the result for the
radiative corrections can be made to be the same. Therefore the
definition of the transverse delta function does not have to be
unique, but it is preferable to keep the symmetry of the theory in
dimensional regularization.

Diagrams (a) and (b) in Fig.~\ref{corad} yield
\begin{eqnarray} \label{mco}
M_a = M_b &=& -2ig^2 C_F \frac{\FMslash{n}}{2} \int \frac{d^D
  l}{(2\pi)^D} \frac{n\cdot (l+p)}{l^2 (l+p)^2 n\cdot l} \Bigl[
\delta  (\omega -\omega^{\prime}) \delta^{(D-2)} ({\bf k}_{\perp}
  +{\bf p}_{\perp} )   \nonumber \\  
&& -\delta
  (\omega -\omega^{\prime}-2 n\cdot l) \delta^{(D-2)} ({\bf l}_{\perp}
  +{\bf k}_{\perp} +{\bf p}_{\perp} ) \Bigr],
\end{eqnarray}
with $\omega^{\prime} = n\cdot p +n\cdot p^{\prime}$. The
prescription for the poles is such that $+i0$ is added to each
factor in the denominator.  
The first integral in Eq.~({\ref{mco}) is simple because the delta
  functions do not involve the loop momentum, and it is given as
\begin{eqnarray} \label{i1}
I_1 &=&  \int \frac{d^D   l}{(2\pi)^D} \frac{n\cdot (l+p)}{l^2 (l+p)^2
  n\cdot l} \delta  (\omega -\omega^{\prime}) \delta^{(D-2)} ({\bf
  k}_{\perp} ) \nonumber \\
&=& -\frac{i}{16\pi^2} \Bigl( \frac{-p^2}{\mu^2}
  \Bigr)^{-\eu} \frac{1}{\eu} \Bigl(- \frac{1}{\ei} -1\Bigr)\delta
  (\omega -\omega^{\prime}) \delta^{(2-2\epsilon)} ({\bf 
  k}_{\perp} ).  
\end{eqnarray}
Note that there appear infrared poles of $\ei$ even though nonzero
$p^2$ is introduced as the infrared cutoff. 
The second integral is more complicated and it is given by
\begin{equation}
I_2 =- \int \frac{d^D   l}{(2\pi)^D} \frac{n\cdot (l+p)}{l^2 (l+p)^2
  n\cdot l} \delta
  (\omega -\omega^{\prime}-2 n\cdot l) \delta^{(D-2)} ({\bf l}_{\perp}
  +{\bf k}_{\perp}+{\bf p}_{\perp}  ).
\end{equation}
After integrating the delta functions, the remaining integral
can be performed using the contour integral in the complex
$\overline{n} \cdot l$ plane. Depending on the position of the poles,
the integral consists of two parts, each of which is proportional to
$\theta (\omega^{\prime} -\omega) \theta (\omega)$ or $\theta (\omega
- \omega^{\prime}) \theta (-\omega)$. The first one corresponds to the
contribution from a quark, while the second one is the antiquark
contribution. Extracting only the contribution from a quark, the
integral becomes
\begin{equation}\label{i2}
I_2 = \frac{i}{16\pi^3} \frac{\omega}{\omega^{\prime}}
  \frac{\theta(\omega^{\prime} -\omega)\theta
  (\omega)}{\omega^{\prime} -\omega} \Bigl[\Bigl({\bf k}_{\perp}
  +\frac{\omega}{\omega^{\prime}} {\bf p}_{\perp} \Bigr)^2
  +\frac{\omega (\omega^{\prime}-\omega)}{(\omega^{\prime})^2}
  \Bigl( \frac{-p^2}{\mu^2} \Bigr) \Bigr]^{-1}. 
\end{equation}
 
We can interpret the characteristics of the integrals $I_1$ and $I_2$ by
tracing where each term arises. In $I_1$, the delta functions do not
include the loop momentum. It means that $I_1$ 
comes from the combinations $\chi_n = W_n^{\dagger} \xi_n$ or
$\overline{\chi}_n =\overline{\xi}W_n$ by extracting a collinear gluon
and attaching it 
to the fermion on the same side with respect to the delta
functions. On the other hand, $I_2$ comes from the configuration in
which a gluon is attached to the fermion on the opposite side with
respect to the delta functions. In other words, $I_1$ corresponds to
the virtual correction and $I_2$ to the real gluon emission in the
full theory. In inclusive deep inelastic scattering, the divergence of
the form $1/(\eu\ei)$ cancels between these two
contributions \cite{Chay:2005rz}. However, comparing Eqs.~(\ref{i1})
and (\ref{i2}), the divergence does not seem to cancel. In the
on-shell limit
$p^2 =0$, as $\omega^{\prime} \rightarrow\omega$ and ${\bf
  k}_{\perp}\rightarrow -{\bf p}_{\perp}$, $I_2$ diverges as
\begin{equation} \label{nlimit}
\frac{1}{\omega^{\prime} -\omega} \frac{1}{({\bf k}_{\perp}+ {\bf
    p}_{\perp})^2}, 
\end{equation}
but it does not cancel the divergence of the virtual gluon
correction. Therefore it was claimed that if all the collinear
particles were on the lightcone and the dimensional regularization was
used, the divergence from the virtual and the real gluon
corrections would not cancel unless an improved regularization
technique was employed. 
  
Here an improved renormalization technique is suggested using
dimensional regularization, in which the divergences between the
virtual and the real gluon corrections cancel. The point is that care
must be taken in extracting a delta function in dimensional 
regularization. In Eq.~(\ref{i2}), the interesting limit is
$\omega^{\prime} \rightarrow \omega$ and ${\bf k}_{\perp} \rightarrow
-{\bf p}_{\perp}$. The naive limit in Eq.~(\ref{nlimit}) does not
extract the correct singular behavior in dimensional
regularization. Note that there is another parameter $-p^2$ in
Eq.~(\ref{i2}), which is introduced as an infrared cutoff. This can be
made arbitrarily small before any physical limits are taken. Therefore 
the basic idea is that we have to take the limit $p^2 \rightarrow 0$
first, and then take other limits. The dimensional regularization
changes the short- and long-distance behavior of the loop integral,
and the transverse momentum and 
the $n$-component of the momentum are closely related to each other as
a result. By putting $p^2 =0$ before any dependence on the dimension
$D=4-2\epsilon$ is computed, it ruins this relation. In this respect,
consider the limit 
\begin{equation}
\lim_{a\rightarrow 0} \frac{1}{{\bf k}_{\perp}^2 +a}. 
\end{equation}
As it is, it diverges when ${\bf k}_{\perp} \rightarrow 0$. However,
we can consider the function
\begin{equation}
f(a, {\bf k}_{\perp}) = \frac{a^{\delta}}{{\bf k}_{\perp}^2 +a} 
\end{equation}
and take the limit $a\rightarrow 0$ such that $f(a,{\bf k}_{\perp})$
can be defined as a distribution function. As ${\bf k}_{\perp}
\rightarrow 0$, $f(a,{\bf k}_{\perp}) \sim a^{-1+\delta}$ and for
${\bf k}_{\perp} \ne 0$, $f (a,{\bf k}_{\perp}) \sim a^{\delta}$. For
$0<\delta <1$, $f(a,{\bf k}_{\perp})$ 
can be regarded as a representation of the delta function in the limit
$a\rightarrow 0$, that is,   
\begin{equation}
\lim_{a\rightarrow 0} \frac{a^{\delta}}{{\bf k}_{\perp}^2 +a} = A
\delta^{(D-2)} ({\bf k}_{\perp}). 
\end{equation}
The normalization constant $A$ is independent of $a$, and is
determined by the integral 
\begin{equation} \label{inta}
\int\frac{d^{D-2} {\bf k}_{\perp}}{(2\pi)^{D-2}}
\frac{a^{\delta}}{{\bf k}_{\perp}^2 +a} = \frac{\Gamma
  (\epsilon)}{(4\pi)^{1-\epsilon}} a^{\delta -\epsilon}
=\frac{A}{(2\pi)^{2- 2 \epsilon}}.  
\end{equation}
If we choose $\delta = \epsilon$, $A= \pi^{1-\epsilon}
\Gamma(\epsilon)$. Therefore as $a\rightarrow 0$, we can write
\begin{equation}\label{lima}
\frac{1}{{\bf k}_{\perp}^2 +a} \rightarrow \pi^{1-\epsilon} \Gamma
(\epsilon) a^{-\epsilon} \delta^{(2-2\epsilon)} ({\bf k}_{\perp}).  
\end{equation}
This holds for $0<\epsilon <1$, and it can be analytically continued
for other values of $\epsilon$. The pole is of the ultraviolet origin,
as can be seen from Eq.~(\ref{inta}). When the representation of the
delta function is used, it can be clearly seen that the momentum
cutoff $|{\bf k}_{\perp}|<Q$ \cite{Lepage:1980fj} cannot be used since
the representation of a delta function is obtained only when all the
range of the transverse momentum is included. Otherwise, 
there will be no cancellation of divergences, and the anomalous
dimension will depend on the cutoff. Actually the integration should
be performed for all ${\bf   k}_{\perp}$ in SCET, since the upper
bound for the effective theory is pulled up to infinity, and the
modification of the high-energy behavior is implemented in the Wilson
coefficients.

If Eq.~(\ref{lima}) is applied to $I_2$, it becomes
\begin{eqnarray}
I_2 &=& \frac{i}{16\pi^2} \frac{\omega}{\omega^{\prime}}
\Bigl(\frac{-p^2 \omega}{\mu^2 \omega^{\prime 2}}\Bigr)^{-\epsilon}
(\omega^{\prime} 
-\omega)^{-1-\epsilon} \theta (\omega^{\prime} -\omega) \theta
(\omega) \Gamma(\epsilon) \delta^{(2-2\epsilon)} \Bigl({\bf
  k}_{\perp}+ \frac{\omega}{\omega^{\prime}} {\bf p}_{\perp} \Bigr) \\  
&=& \frac{i}{16\pi^2} \Bigl(\frac{-p^2}{\mu^2}\Bigr)^{-\epsilon} \Bigl[
-\frac{1}{\eu\ei} \delta (\omega -\omega^{\prime}) +\frac{1}{\eu}
\frac{\omega}{\omega^{\prime}}
\frac{\theta(\omega^{\prime}-\omega)}{(\omega^{\prime}
  -\omega)_+}\Bigr] \theta (\omega) \delta^{(2-2\epsilon)} \Bigl({\bf
  k}_{\perp}+ \frac{\omega}{\omega^{\prime}} {\bf p}_{\perp}
  \Bigr). \nonumber 
\end{eqnarray}
We can write
\begin{equation} \label{modx}
(\omega^{\prime} -\omega)^{-1-\epsilon} \theta (\omega^{\prime}
  -\omega)\theta (\omega) = (\omega^{\prime})^{-1-\epsilon}
  (1-x)^{-1-\epsilon} \theta (1-x) \theta (x),
\end{equation}
where $x=\omega/\omega^{\prime}$. When we integrate over $x$, the
integral diverges at $x\approx 1$, which is of the infrared
origin. Since the limits of the integral is bounded, there is no
ultraviolet divergence and we obtain the formula
\begin{equation} \label{plusf}
(\omega^{\prime} -\omega)^{-1-\epsilon} \theta (\omega^{\prime}
  -\omega)\theta (\omega) =\Bigl[-\frac{1}{\ei} \delta
    (\omega^{\prime}-\omega) 
  +\frac{\theta( \omega^{\prime}-\omega)}{(\omega^{\prime} -\omega)_+}
  \Bigr] \theta (\omega),
\end{equation}
with the $+$-distribution function. In the above expression, the
divergence of the form $1/\eu\ei$ is cancelled in the sum $I_1 +I_2$,
and the amplitudes are given as 
\begin{equation}
M_a = M_b = \frac{\alpha_s C_F}{4\pi} \frac{\FMslash{n}}{2}
\frac{2}{\eu} \theta (\omega)
\Bigl[ \delta 
(\omega -\omega^{\prime}) \delta^{(2-2\epsilon)} ({\bf
  k}_{\perp})+\frac{\omega}{\omega^{\prime}} \frac{\theta( 
    \omega^{\prime}-\omega)}{(\omega^{\prime} -\omega)_+}
\delta^{(2-2\epsilon)} \Bigl({\bf k}_{\perp}
+\frac{\omega}{\omega^{\prime}} {\bf p}_{\perp} \Bigr)\Bigr].
\end{equation}
In a similar way, $M_c$ can be calculated to give
\begin{equation}
M_c =\frac{\alpha_s C_F}{4\pi} \frac{-2}{\eu} \delta^{(2-2\epsilon)}
    \Bigl({\bf k}_{\perp} +\frac{\omega}{\omega^{\prime}} {\bf
    p}_{\perp} \Bigr) \frac{\omega^{\prime} -\omega}{\omega^{\prime
    2}} \theta (\omega^{\prime} -\omega) \theta (\omega). 
\end{equation}
Adding all the contributions, we have
\begin{eqnarray}
M_a +M_b +M_c &=& \frac{\alpha_s C_F}{2\pi} \frac{\FMslash{n}}{2}
\frac{1}{\eu} \Bigl[ 2 \delta (\omega -\omega^{\prime})
  \delta^{(2-2\epsilon)} ({\bf k}_{\perp}) \nonumber \\
&+&\frac{1+
    (\omega/\omega^{\prime} )^2}{(\omega^{\prime} -\omega)_+} \theta
  (\omega^{\prime} -\omega) \theta (\omega) \delta^{(2-2\epsilon)}
    \Bigl({\bf k}_{\perp} +\frac{\omega}{\omega^{\prime}} {\bf
    p}_{\perp} \Bigr)\Bigr]. 
\end{eqnarray}

Now we have to subtract the zero-bin contribution
\cite{Manohar:2006nz}, which corresponds to the soft limit when
the collinear loop momentum becomes soft, 
i.e., of order $\Lambda$. The collinear contribution from this soft
region should be subtracted since collinear particles are defined to
have large label momentum. This region should be described by soft
particles, not by collinear particles with small label
momenta. The loop integrals in $M_a +M_b +M_c$ were performed naively
including the contribution of the collinear particles with small label
momenta. Therefore this contribution should be subtracted. 
In computing the naive collinear contribution, the loop and the
external momenta scale as
\begin{equation}
l^{\mu} =(n\cdot l, l_{\perp}, \overline{n}\cdot l) \sim (Q, \Lambda,
\Lambda^2 /Q), \ \ p^{\mu} =(n\cdot p, p_{\perp}, \overline{n}\cdot p)
\sim  (Q, \Lambda, \Lambda^2 /Q),
\end{equation}
But in computing the zero-bin contribution, we subtract the
contribution where the largest component $n\cdot l$ becomes $\Lambda$,  
retaining the same scaling for $\overline{n}\cdot l$ and adjusting the
scaling of $l_{\perp}$ such that $n\cdot l \overline{n} \cdot l \sim
l_{\perp}^2$.  Therefore the power
counting of the loop and the external momenta is given by
\begin{equation} \label{pc}
l^{\mu} =(n\cdot l, l_{\perp}, \overline{n}\cdot l) \sim (\Lambda,
\Lambda \sqrt{\Lambda/Q} , \Lambda^2 /Q), \ \ p^{\mu} =(n\cdot p,
p_{\perp}, \overline{n}\cdot p) \sim  (Q, \Lambda, \Lambda^2 /Q). 
\end{equation}
One may wonder why the loop momentum of order $l^2 \sim \Lambda^3 /Q$
is subtracted, hence unphysical. That is exactly the reason why it
should be subtracted from the collinear contribution and replaced by
the soft contribution in which all the components of the momentum are
of order $\Lambda$. We 
can also choose the power counting as $n\cdot l \sim \Lambda$, 
while others are fixed such that $l_{\perp}^2 \gg n\cdot
l \overline{n}\cdot l$, which is called the Glauber region. But it
does not alter the result of the zero-bin contribution. The point of
the zero-bin subtraction is to subtract the part in which energetic
collinear particles become soft, and the criterion is the size of the
label momentum.

According to the power counting in Eq.~(\ref{pc}), the zero-bin
contribution from the naive collinear integrals $M_a$ and $M_b$ are
given as 
\begin{eqnarray}
M_a^0 = M_b^0 &=& -2ig^2 C_F \frac{\FMslash{n}}{2} \int \frac{d^D
  l}{(2\pi)^D} \frac{1}{l^2 (\overline{n} \cdot l +p^2/n\cdot p)
  n\cdot l} \nonumber \\
&&\times \Bigl[ \delta (\omega -\omega^{\prime}) \delta^{(D-2)} ({\bf 
  k}_{\perp} +{\bf p}_{\perp}) -\delta (\omega -\omega^{\prime}
  -2n\cdot l)   \delta^{(D-2)} (\bf{k}_{\perp}+{\bf p}_{\perp})
  \Bigr],  
\end{eqnarray}
and the zero-bin contribution from $M_c$ vanishes. 
The first integral is given by
\begin{eqnarray}
J_1 &=& \int \frac{d^D   l}{(2\pi)^D}  \frac{1}{l^2 (\overline{n}
  \cdot l +p^2/n\cdot p)   n\cdot l} 
 \delta (\omega -\omega^{\prime}) \delta^{(D-2)} ({\bf
  k}_{\perp}+{\bf p}_{\perp}) \nonumber \\
&=&-\frac{i}{16\pi^2} \Bigl( \frac{-p^2}{\mu n\cdot p} \Bigr)^{-\eu}
  \Bigl( \frac{1}{\eu} -\frac{1}{\ei} \Bigr) \frac{1}{\eu}  \delta
  (\omega -\omega^{\prime}) \delta^{(2-2\epsilon)} ({\bf
  k}_{\perp}+{\bf p}_{\perp}).  
\end{eqnarray}
The second integral $J_2$ can be computed by integrating over the delta
function first, doing the contour integral in the complex
$\overline{n}\cdot l$ plane, and then finally integrating over
$l_{\perp}$ using dimensional regularization. $J_2$ is given as
\begin{eqnarray}
J_2 &=& -\int \frac{d^D l}{(2\pi)^D} \frac{1}{l^2 (\overline{n} \cdot l
  +p^2/n\cdot p)   n\cdot l} \delta (\omega -\omega^{\prime} -2n\cdot l)
  \delta^{(D-2)} (\bf{k}_{\perp}+{\bf p}_{\perp})  \\
&=& \frac{i}{16\pi^2} \frac{1}{\eu} \frac{\theta (\omega^{\prime}
  -\omega)}{\omega^{\prime} -\omega} \Bigl(\frac{(\omega^{\prime}
  -\omega) (-p^2)}{n\cdot p \mu} \Bigr)^{-\eu}  \delta^{(2-2\epsilon)}
  (\bf{k}_{\perp}+{\bf p}_{\perp}) \nonumber \\
&=& \frac{i}{16\pi^2} \Bigl(\frac{-p^2}{ n\cdot p \mu} \Bigr)^{-\eu}
  \frac{1}{\eu} 
  \Bigl[   \Bigl( \frac{1}{\eu} -\frac{1}{\ei}\Bigr) \delta (\omega
  -\omega^{\prime}) +\frac{\theta (\omega^{\prime}
  -\omega)}{(\omega^{\prime} -\omega)_+} \Bigr]\delta^{(2-2\epsilon)}
  (\bf{k}_{\perp}+{\bf p}_{\perp}), \nonumber 
\end{eqnarray}
where the following formula is used:
\begin{equation} \label{noplusf}
\theta (\omega^{\prime} -\omega) (\omega^{\prime}
-\omega)^{-1-\epsilon}  = \Bigl( \frac{1}{\eu} -\frac{1}{\ei}\Bigr)
\delta (\omega^{\prime} -\omega) + \frac{\theta (\omega^{\prime}
  -\omega)}{(\omega^{\prime} -\omega)_+}.  
\end{equation}
The difference between this equation and Eq.~(\ref{plusf}) is that
there is no lower bound. Therefore
Eq.~(\ref{noplusf}) has both infrared and ultraviolet singularity.
In the sum $J_1 +J_2$, the $\delta (\omega^{\prime} -\omega)$ term
cancels and the zero-bin contribution is given by
\begin{equation}
M_a^0 =M_b^0 = \frac{\alpha_s C_F}{2\pi} \frac{\FMslash{n}}{2}
\frac{2}{\eu} \frac{\theta (\omega^{\prime} -\omega)}{(\omega^{\prime}
  -\omega)_+} \delta^{(2-2\epsilon)} ({\bf k}_{\perp}+{\bf p}_{\perp}).
\end{equation}

Subtracting the zero-bin contribution, the total collinear
contribution with the wavefunction renormalization is given by
\begin{eqnarray} \label{tmdco0}
&&M_a +M_b +M_c -(M_a^0 +M_b^0) +\mathrm{w.f.}  \nonumber \\
& =& \frac{\alpha_s
  C_F}{2\pi} \frac{\FMslash{n}}{2} \frac{2}{\eu} \Bigl[ \Bigl(
    \frac{3}{2}   \delta (\omega^{\prime} -\omega)
    -\frac{2}{(\omega^{\prime}  -\omega)_+ } \theta (\omega^{\prime}
    -\omega) \theta (\omega)\Bigr)
  \delta^{(2-2\epsilon)} ({\bf k}_{\perp}+{\bf p}_{\perp}) \nonumber \\
&&+  \frac{1+(\omega/\omega^{\prime})^2}{(\omega^{\prime}
    -\omega)_+}\theta (\omega^{\prime}     -\omega) \theta (\omega) 
 \delta^{(2-2\epsilon)} ({\bf k}_{\perp}+
  \frac{\omega}{\omega^{\prime}} {\bf p}_{\perp}) \Bigr],
\end{eqnarray}
where $\theta (\omega)$ in the second term is inserted for physical
reasons. 
For comparison, the corresponding radiative correction for the
collinear operator giving the integrated parton distribution function
is given as \cite{Chay:2005rz}
\begin{equation}
\frac{\alpha_s  C_F}{2\pi} \frac{\FMslash{n}}{2} \frac{2}{\eu} \Bigl[
    \frac{3}{2}   \delta (\omega^{\prime} -\omega)
    -\frac{2}{(\omega^{\prime}  -\omega)_+ } \theta (\omega^{\prime}
    -\omega) \theta (\omega) 
+  \frac{1+(\omega/\omega^{\prime})^2}{(\omega^{\prime}
    -\omega)_+}\theta (\omega^{\prime}     -\omega) \theta (\omega) 
\Bigr].
\end{equation}
This is the result from Eq.~(\ref{tmdco0}) without the delta functions
of the transverse momentum, or by integrating over the transverse
momentum. However, note that the
radiative correction for the collinear TMD operator is not just
proportional to $\delta^{(2-2\epsilon)} ({\bf k}_{\perp} +{\bf
  p}_{\perp})$, but it also depends on $\delta^{(2-2\epsilon)} ({\bf 
  k}_{\perp} + (\omega^{\prime}/\omega){\bf   p}_{\perp})$.
This gives a nontrivial difference in the renormalization group
behavior between the TMD and the integrated PDF.

\subsection{One loop correction to the soft Wilson line}

We now compute the radiative corrections for the TMD soft Wilson line
operator 
\begin{equation}
S(\eta, {\bf s}_{\perp}) = \frac{1}{N}
\mathrm{tr} \, \tilde{Y}^{\dagger}_{\bar{n}} Y_n \delta (\eta +n\cdot
i\partial  )
\delta^{(2)} ({\bf s}_{\perp} + i\nabla_{\perp})  Y_n^{\dagger}
Y_{\bar{n}} .
\end{equation}
The Feynman rule for the operator $S$ with two gluons is given in
Fig.~\ref{sfeyn}, and the Feynman diagram for the radiative correction
at one loop is shown in 
Fig.~\ref{sloop}. The radiative correction is written as
\begin{eqnarray} \label{tmdsfr}
I(\eta, {\bf s}_{\perp} ) &=& -2ig^2 C_F \int \frac{d^D l}{(2\pi)^D}
\Bigl[ \frac{2 \delta (\eta) \delta^{(D-2)} ({\bf s}_{\perp})}{ l^2
    (\overline{n} \cdot l -\lambda_1 ) (n\cdot l -\lambda_2)} \\
&+& \frac{\delta (\eta +n\cdot l) \delta^{(D-2)} ({\bf s}_{\perp}+{\bf 
      l}_{\perp})}{(l^2 +i0) (\overline{n}\cdot l -\lambda_1 +i0)}
   \frac{1}{-n\cdot l +\lambda_2}  
+ \frac{\delta (\eta -n\cdot l) \delta^{(D-2)} ({\bf s}_{\perp} -{\bf
      l}_{\perp})}{(l^2 +i0)(-\overline{n} \cdot l -\lambda_1 +i0)}
  \frac{1}{n\cdot l +\lambda_2} \Bigr], \nonumber 
\end{eqnarray}
where $\lambda_1$ and $\lambda_2$ are the infrared cutoffs. And the
necessary $i0$ prescriptions for the contour integral in the
complex $\overline{n}\cdot l$ plane are shown.

\begin{figure}[t] 
\begin{center}
\includegraphics[height=2.5cm]{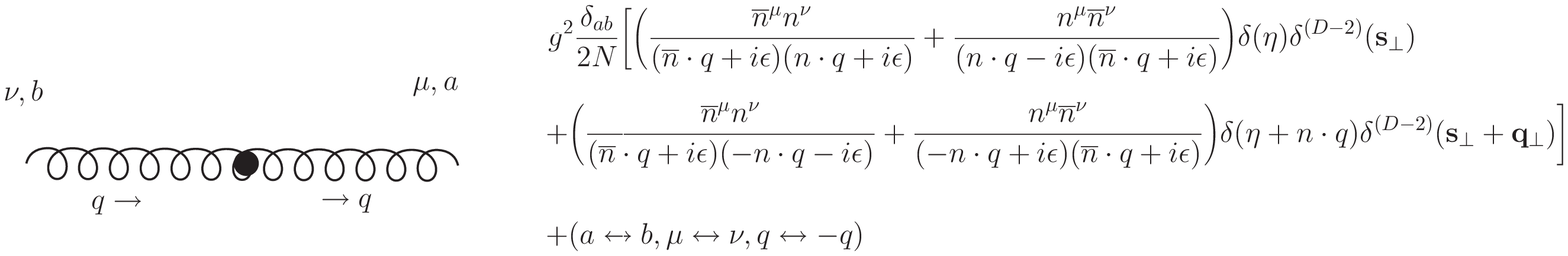}
\end{center}
\vspace{-0.5cm}
\caption{\baselineskip 3.0ex
Feynman rules for the soft Wilson line operator $S(\eta, {\bf
  s}_{\perp})$ with two soft gluons. 
\label{sfeyn} }   
\end{figure}

\begin{figure}[b] 
\begin{center}
\includegraphics[height=3.0cm]{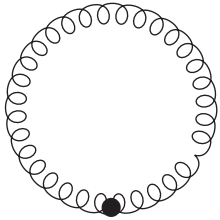}
\end{center}
\vspace{-0.5cm}
\caption{\baselineskip 3.0ex
Feynman diagram of one-loop correction for the TMD soft Wilson
line.\label{sloop} }    
\end{figure}

The first integral is given by
\begin{eqnarray}
S_1 &=&-4 ig^2 C_F \int \frac{d^D l}{(2\pi)^D} \frac{\delta (\eta)
    \delta^{(D-2)} ({\bf s}_{\perp})}{ l^2 
    (\overline{n} \cdot l -\lambda_1 ) (n\cdot l -\lambda_2)}
    \nonumber \\ 
&=&    -\frac{\alpha_s C_F}{2\pi} \frac{2}{\eu^2}
    \Bigl(\frac{\lambda_1     \lambda_2}{\mu^2} \Bigr)^{-\epsilon}
    \delta (\eta)     \delta^{(2-2\epsilon)} ({\bf s}_{\perp}). 
\end{eqnarray}
In the second and the third integrals in Eq.~(\ref{tmdsfr}), we
perform the contour 
integration in the complex $\overline{n}\cdot l$ place after
integrating over the delta functions. The second and the third
integrals are given by 
\begin{equation}
S_2 =S_3 = \frac{g^2 C_F}{8\pi^3} \frac{\theta (\eta)}{\eta+\lambda_2}
\frac{1}{{\bf s}_{\perp}^2 +\eta \lambda_1}. 
\end{equation}
We apply the idea of the representation of the delta function again to
have
\begin{equation}
\lim_{\lambda_1 \rightarrow 0} \frac{1}{{\bf s}_{\perp}^2 +\eta
  \lambda_1} = \frac{\pi^{1-\epsilon}}{\eu} (\eta
  \lambda_1)^{-\epsilon}   \delta^{(2-2\epsilon)} ({\bf s}_{\perp}),
\end{equation}
and the integral becomes
\begin{equation}
S_2 = \frac{\alpha_s C_F}{2\pi} \frac{\theta (\eta)}{\eu} \frac{(\eta
  \lambda_1)^{-\epsilon}}{\eta +\lambda_2} \delta^{(2-2\epsilon)} ({\bf
  s}_{\perp}) =  \frac{\alpha_s C_F}{2\pi}  \Bigl[
  \Bigl(\frac{\lambda_1 \lambda_2}{\mu^2}\Bigr)^{-\epsilon} 
  \frac{1}{\eu^2} +\frac{1}{\eu} \frac{\theta (\eta)}{\eta_+} \Bigr]
  \delta^{(2-2\epsilon)} ({\bf s}_{\perp}).   
\end{equation}
Here the plus distribution function is also used. But note that all
the poles in $1/\epsilon$ are of ultraviolet origin since the infrared
cutoffs $\lambda_1$ and $\lambda_2$ are used. Adding all these, the
radiative correction for the soft Wilson line at one loop is given by
\begin{equation} \label{softl}
I(\eta,{\bf s}_{\perp}) =\frac{\alpha_s C_F}{\pi} \frac{1}{\eu}
\frac{\theta (\eta)}{\eta_+}  \delta^{(2-2\epsilon)} ({\bf s}_{\perp}). 
\end{equation}
The only difference, compared with the integrated soft Wilson line
\cite{Chay:2004zn}, is the presence of $\delta^{(2-2\epsilon)} ({\bf
  s}_{\perp})$. Therefore the anomalous dimension is the same as the
integrated soft Wilson line except the delta function $
\delta^{(2-2\epsilon)} ({\bf s}_{\perp})$, that is, the TMD soft
Wilson line does not scale with respect to the transverse momentum.

\section{Renormalization group behavior of the TMD
  operators\label{renorm}} 
With the radiative corrections, the relations between the bare
operators and the renormalized operators can be written, in general,
as 
\begin{eqnarray}
\mathcal{O}_c^B (\omega, {\bf k}_{\perp}; {\bf p}_{\perp}) &=& \int
d\omega^{\prime} 
d^{D-2} {\bf k}^{\prime}_{\perp} Z_C(\omega, \omega^{\prime}, {\bf
  k}_{\perp}, {\bf k}^{\prime}_{\perp}; {\bf p}_{\perp})
\mathcal{O}_c^R (\omega^{\prime}, {\bf k}_{\perp}^{\prime}; {\bf
  p}_{\perp}),  \nonumber \\ 
S^B (\eta, {\bf s}_{\perp}) &=& \int d\eta^{\prime}
d^{D-2} {\bf s}^{\prime}_{\perp} Z_S(\eta, \eta^{\prime}, {\bf
  s}_{\perp}, {\bf s}^{\prime}_{\perp}) S^R (\eta^{\prime}, {\bf
  s}^{\prime}_{\perp}) ,  
\end{eqnarray}
where the counterterm kernels $Z_C(\omega, \omega^{\prime}, {\bf
  k}_{\perp}, {\bf k}^{\prime}_{\perp};{\bf p}_{\perp})$ and
  $Z_S(\eta, \eta^{\prime}, {\bf   s}_{\perp}, {\bf
  s}^{\prime}_{\perp})$ can be read off from Eqs.~(\ref{tmdco0}) and
  (\ref{softl}), and are given as 
\begin{eqnarray} \label{zczs}
Z_C (\omega, \omega^{\prime}, {\bf   k}_{\perp}, {\bf
  k}^{\prime}_{\perp}; {\bf p}_{\perp}) &=&  \delta
  (\omega^{\prime} -\omega) \delta^{(2-2\epsilon)} ({\bf
  k}_{\perp} -{\bf k}_{\perp}^{\prime}) \nonumber \\
&+& \frac{\alpha_s
  C_F}{2\pi} \frac{\FMslash{n}}{2} \frac{2}{\eu}  \Bigl[\Bigl(
    \frac{3}{2}   \delta (\omega^{\prime} -\omega)
    -\frac{2 \theta (\omega^{\prime} -\omega) \theta (\omega)
  }{(\omega^{\prime}  -\omega)_+ }  \Bigr)
  \delta^{(2-2\epsilon)} ({\bf k}_{\perp}-{\bf k}_{\perp}^{\prime} )
  \nonumber   \\ 
&+&  \frac{1+(\omega/\omega^{\prime})^2}{(\omega^{\prime}
    -\omega)_+}\theta (\omega^{\prime}     -\omega) \theta (\omega) 
 \delta^{(2-2\epsilon)} ({\bf k}_{\perp}-{\bf k}_{\perp}^{\prime} -
  \frac{\omega^{\prime} -\omega}{\omega^{\prime}} {\bf p}_{\perp})
  \Bigr],   \nonumber \\
Z_S(\eta, \eta^{\prime}, {\bf   s}_{\perp}, {\bf s}^{\prime}_{\perp})
  &=&  \delta^{(2-2\epsilon)} ({\bf s}_{\perp} -{\bf
  s}_{\perp}^{\prime}) 
  \Bigl[ \delta (\eta -\eta^{\prime}) +\frac{\alpha_s C_F}{\pi \eu}
  \frac{\theta (\eta -\eta^{\prime})}{(\eta -\eta^{\prime})_+}\Bigr].  
\end{eqnarray}
Here in $Z_C(\omega, \omega^{\prime}, {\bf   k}_{\perp}, {\bf
  k}^{\prime}_{\perp};{\bf p}_{\perp})$, the external transverse
 momentum ${\bf p}_{\perp}$ of the quark is introduced for
 convenience, but it should be understood that it appears in the
  matrix element of the  operator $\mathcal{O}_c$.    
The renormalization group equations for the renormalized operators are
written as
\begin{eqnarray}\label{rg}
\mu \frac{d}{d\mu} \mathcal{O}_c^R (\omega, {\bf k}_{\perp};{\bf
  p}_{\perp}) &=& -\int 
d\omega^{\prime} d^2 {\bf k}_{\perp}^{\prime} \gamma_C (\omega,
\omega^{\prime}, {\bf   k}_{\perp}, {\bf   k}^{\prime}_{\perp};{\bf
  p}_{\perp}) \mathcal{O}_c^R (\omega^{\prime}, {\bf
  k}_{\perp}^{\prime}), \nonumber \\
\mu \frac{d}{d\mu} S^R (\eta, {\bf s}_{\perp}) &=& -\int
d\eta^{\prime} d^2 {\bf s}_{\perp}^{\prime} \gamma_S (\eta,
  \eta^{\prime}, {\bf   s}_{\perp}, {\bf s}^{\prime}_{\perp})  S^R
  (\eta^{\prime}, {\bf   s}_{\perp}^{\prime}),  
\end{eqnarray}
where the anomalous dimension kernels are given by
\begin{eqnarray} \label{anomcs}
\gamma_C (\omega, \omega^{\prime}, {\bf   k}_{\perp}, {\bf
  k}^{\prime}_{\perp}; {\bf p}_{\perp}) &=& -\frac{\alpha_s C_F}{\pi}
  \Bigl[ \Bigl(  \frac{3}{2}  \delta (\omega -\omega^{\prime}) 
- \frac{2 \theta (\omega^{\prime} -\omega) \theta
  (\omega)}{(\omega^{\prime} -\omega)_+} \Bigr) \delta^{(2)} ({\bf
  k}_{\perp}   -{\bf k}_{\perp}^{\prime})  \nonumber  \\
&+& \frac{1+(\omega/\omega^{\prime})^2}{(\omega^{\prime}
    -\omega)_+}\theta (\omega^{\prime}     -\omega) \theta (\omega) 
 \delta^{(2)} ({\bf k}_{\perp}-{\bf k}^{\prime}_{\perp} -
  \frac{\omega^{\prime} -\omega}{\omega^{\prime}}
 {\bf p}_{\perp}) \Bigr], \nonumber \\ 
\gamma_S (\eta, \eta^{\prime},
{\bf   s}_{\perp}, {\bf s}^{\prime}_{\perp}) &=& -2 \frac{\alpha_s
  C_F}{\pi} \delta^{(2)} ({\bf s}_{\perp} -{\bf s}_{\perp}^{\prime})
  \frac{\theta (\eta -\eta^{\prime})}{(\eta -\eta^{\prime})_+}.  
\end{eqnarray}
Note that the second term in $\gamma_C$ in Eq.~(\ref{anomcs}), which
corresponds to the zero-bin subtraction has the same form in
$\gamma_S$ with the opposite sign. It confirms the nature of the
zero-bin subtraction, which is computed in the soft limit.
This pattern also appears in integrated PDF \cite{Chay:2005rz}. 

It is useful to express the renormalization group equation for
$\mathcal{O}_c$ in terms of the dimensionless variables $y$ and $z$
defined as $\omega =2yn\cdot P_{\bar{n}}$, $\omega^{\prime}= 2z n\cdot
P_{\bar{n}}$, where $0<y,z<1$. It is given as
\begin{equation} \label{tmdren}
\mu \frac{d}{d\mu} \mathcal{O}_c (y,{\bf k}_{\perp}; {\bf p}_{\perp})
= - \int_y^1 dz \int d^2 {\bf k}_{\perp}^{\prime}\gamma_C (y,z,{\bf
  k}_{\perp}, {\bf k}_{\perp}^{\prime};{\bf p}_{\perp}) \mathcal{O}_c
(z,{\bf k}_{\perp}^{\prime};{\bf p}_{\perp}),\
\end{equation}
where
\begin{eqnarray}
\gamma_C (y,z,{\bf  k}_{\perp}, {\bf k}_{\perp}^{\prime};{\bf
  p}_{\perp}) &=& -\frac{\alpha_s C_F}{\pi} 
\Bigl[ \Bigl( \frac{3}{2} \delta (y-z)  -
  \frac{2}{(z-y)_+} \Bigr)  \delta^{(2)} ({\bf  k}_{\perp}   -{\bf
    k}_{\perp}^{\prime})   \nonumber \\ 
&+& \frac{1+(y/z)^2}{(z-y)_+} \delta^{(2)} \Bigl({\bf
    k}_{\perp} -{\bf k}_{\perp}^{\prime} - \frac{z-y}{z} {\bf
    p}_{\perp} \Bigr)
  \Bigr].
\end{eqnarray}
In terms of $x=y/z$, Eq.~(\ref{tmdren}) becomes   
\begin{eqnarray} \label{tmdx}
\mu \frac{d}{d\mu} \mathcal{O}_c (y,{\bf k}_{\perp}; {\bf p}_{\perp}) 
&=& \frac{\alpha_s C_F}{\pi} \int d^2 {\bf k}^{\prime}_{\perp}
\int_y^1 \frac{dx}{x} \Bigl[ \Bigl( \frac{3}{2} \delta ( 1-x)  -
  \frac{2}{(1-x)_+} \Bigr)  \delta^{(2)} ({\bf  k}_{\perp}   -{\bf
    k}_{\perp}^{\prime})   \nonumber \\ 
&+& \frac{1+x^2}{(1-x)_+} \delta^{(2)} ({\bf
    k}_{\perp} -{\bf k}_{\perp}^{\prime} -(1-x) {\bf p}_{\perp})
  \Bigr] \mathcal{O}_c (y/x, {\bf k}_{\perp}^{\prime}; {\bf
  p}_{\perp}).
\end{eqnarray}
Note that, from Eq.~(\ref{gk}), the matrix element of the collinear
operator $\mathcal{O}_c$ is given by
\begin{equation}
g(y,-{\bf k}_{\perp}) =\langle P_{\bar{n}}| \mathcal{O}_c (\omega
=2yn\cdot P_{\bar{n}}, {\bf k}_{\perp})|P_{\bar{n}}\rangle. 
\end{equation}
Then Eq.~(\ref{tmdren}) can be expressed in impact parameter
representation, or its Fourier transform as
\begin{eqnarray} \label{gb}
\mu \frac{d}{d\mu}   \tilde{g} (y,{\bf b}_{\perp}) &=& \mu
  \frac{d}{d\mu} \int d^2 {\bf k}_{\perp} e^{i {\bf b}_{\perp} 
  \cdot {\bf k}_{\perp}} g (y,-{\bf k}_{\perp})  \\
&=& \frac{\alpha_s C_F}{\pi} \int_y^1 dz \Bigl[ \frac{3}{2} \delta
  (y-z) +\frac{-2 +(1+(y/z)^2)e^{i (1-y/z) {\bf b}_{\perp} \cdot {\bf
  p}_{\perp}}}{(z-y)_+} \Bigr] \tilde{g} (z,{\bf
  b}_{\perp}), \nonumber 
\end{eqnarray}
where the sign of the exponent becomes opposite compared to
Eq.~(\ref{impar}) since the 4-vector notation is changed to the
3-vector notation here. If we put ${\bf b}_{\perp} =0$ in
Eq.~(\ref{gb}), both sides  become the integrated quantities over the 
transverse momentum ${\bf k}_{\perp}$, and the corresponding
renormalization group equation is given by  
\begin{equation} \label{intpdf}
\mu \frac{d}{d\mu} \int d^2 {\bf k}_{\perp} g (y,{\bf
  k}_{\perp})
=  \frac{\alpha_s }{\pi} \int_y^1 \frac{dx}{x} \Bigl( P_{qq} (x)
  -\frac{2C_F}{(1-x)_+}  \Bigr) 
\int d^2 {\bf k}_{\perp}  g (y/x, {\bf k}_{\perp}) ,   
\end{equation}
where $P_{qq}(x)$ is the quark splitting function given by 
\begin{equation}
P_{qq} (x) = C_F \Bigl[ \frac{3}{2} \delta (1-x)
  +\frac{1+x^2}{(1-x)_+} \Bigr].
\end{equation}
An extra term in Eq.~(\ref{intpdf}) other than $P_{qq} (x)$ comes from
the zero-bin subtraction. Eq.~(\ref{intpdf}) is exactly the same
renormalization group equation for the integrated collinear operator
in Ref.~\cite{Chay:2005rz}.

The renormalization group equation for the TMD soft Wilson line after
integrating over ${\bf s}_{\perp}^{\prime}$ in Eq.~(\ref{rg}) is
written as
\begin{equation}
\mu \frac{d}{d\mu} S (\eta, {\bf s}_{\perp}) = \frac{2\alpha_s
  C_F}{\pi} \int d\eta^{\prime}  \frac{\theta (\eta
  -\eta^{\prime})}{(\eta -\eta^{\prime})_+} 
S (\eta^{\prime}, {\bf   s}_{\perp}). 
\end{equation}
Since the anomalous dimension for the TMD soft Wilson line is
proportional to $\delta ({\bf s}_{\perp}-{\bf s}_{\perp}^{\prime})$,
the transverse-momentum-dependent part is trivial and the TMD soft
Wilson line does not scale with respect to the transverse momentum. As
a result, the renormalization group equation for the TMD soft Wilson
line is the same as the integrated soft Wilson line
\cite{Chay:2004zn}.

Besides the renormalization group equation,
the radiative corrections themselves, when integrated over the
transverse momentum, give the same result for the radiative correction
of the integrated collinear operator. If the counterterm kernels in
Eq.~(\ref{zczs}) is integrated over the transverse momentum, the
results correspond to the counterterm kernels for the integrated
collinear and soft Wilson line operators. Therefore the relation
between the TMD (unintegrated) and integrated PDF
can be clearly seen in this formulation. That is, the
integrated PDF and its renormalization group behavior are obtained by
integrating the TMD PDF and its renormalization group equation over
the transverse momentum. We can see this relation explicitly. From
Eq.~(\ref{fkappa}), the renormalization group equation for
$f(y,{\bf p}_{\perp}, \kappa)$ is written as
\begin{eqnarray}
\mu \frac{d}{d\mu} f(y,{\bf p}_{\perp},\kappa) &=& \int d^2 {\bf
  s}_{\perp}  d^2 {\bf k}_{\perp} \delta^{(2)} ({\bf p}_{\perp} + {\bf
  k}_{\perp} + {\bf s}_{\perp}) \nonumber
\\ 
&\times& \Bigl[  \Bigl( \mu \frac{d}{d\mu} g(y,-{\bf k}_{\perp})\Bigr) 
  K(\kappa,{\bf s}_{\perp}) +g(y,-{\bf k}_{\perp})  \Bigl( \mu
  \frac{d}{d\mu}   K(\kappa,{\bf s}_{\perp}) \Bigr) \Bigr] \nonumber
  \\
&=& -\int d^2 {\bf
  s}_{\perp}  d^2 {\bf k}_{\perp} \delta^{(2)} ({\bf p}_{\perp} + {\bf
  k}_{\perp} + {\bf s}_{\perp}) \int d^2 {\bf   s}_{\perp}^{\prime}
  d^2 {\bf k}_{\perp}^{\prime} dz d\kappa^{\prime} \nonumber \\
&\times& \gamma_f (y,z,{\bf k}_{\perp}, {\bf k}_{\perp}^{\prime};
  \kappa, \kappa^{\prime}, {\bf s}_{\perp}, {\bf s}_{\perp}^{\prime};
  {\bf p}_{\perp}) g(z,-{\bf k}_{\perp}^{\prime}) K(\kappa^{\prime},
  {\bf s}_{\perp}^{\prime}),  
\end{eqnarray}
where
\begin{eqnarray}
\gamma_f (y,z,{\bf k}_{\perp}, {\bf k}_{\perp}^{\prime};
  \kappa, \kappa^{\prime}, {\bf s}_{\perp}, {\bf s}_{\perp}^{\prime};
  {\bf p}_{\perp}) &=& \gamma_C (y,z,{\bf  k}_{\perp}, {\bf
  k}_{\perp}^{\prime};{\bf   p}_{\perp}) \delta(\kappa
  -\kappa^{\prime}) \delta^{(2)} ({\bf s}_{\perp} -{\bf
  s}_{\perp}^{\prime}) \nonumber \\
&+& \gamma_S (\kappa,\kappa^{\prime}, {\bf s}_{\perp}, {\bf
  s}_{\perp}^{\prime}) \delta (y-z) \delta^{(2)} ({\bf k}_{\perp}
  -{\bf k}_{\perp}^{\prime}).  
\end{eqnarray}
This is a complicated integro-differential equation.
But it is more convenient to derive the renormalization group equation
in impact parameter space. From Eq.~(\ref{fimp}), the renormalization
group equation for $\tilde{f}(y,{\bf b}_{\perp},\kappa)$ is given by
\begin{eqnarray} \label{rgf}
\mu \frac{d}{d\mu} \tilde{f} (y,{\bf b}_{\perp},\kappa) &=& \Bigl( \mu  
\frac{d}{d\mu} \tilde{g} (y,{\bf b}_{\perp}) \Bigr) \tilde{K}(\kappa,
-{\bf b}_{\perp})  +\tilde{g} (y,{\bf b}_{\perp})  \Bigl( \mu
\frac{d}{d\mu} \tilde{K}(\kappa, -{\bf b}_{\perp}) \Bigr).
\end{eqnarray}
Each $\tilde{g} (y,{\bf b}_{\perp})$ and $\tilde{K} (\kappa,-{\bf
  b}_{\perp})$ satisfies the renormalization group equation
\begin{eqnarray}
\mu \frac{d}{d\mu} \tilde{g} (y,{\bf b}_{\perp}) &=& \mu
\frac{d}{d\mu} \int d^2 {\bf k}_{\perp} e^{i {\bf b}_{\perp} \cdot
  {\bf k}_{\perp}} g(y,-{\bf k}_{\perp}) \\ 
&=& \frac{\alpha_s C_F}{\pi} \int_y^1 dz \Bigl[ \frac{3}{2} \delta
  (y-z)  +\frac{ -2 + (1+(y/z)^2) e^{i(1-y/z)
    {\bf b}_{\perp} \cdot {\bf p}_{\perp}}}{(z-y)_+} \Bigr]
\tilde{g}(z,  {\bf b}_{\perp} ), \nonumber \\
 \mu \frac{d}{d\mu} \tilde{K} (\kappa,-{\bf b}_{\perp}) &=&
 \frac{2\alpha_s  C_F}{\pi} \int d\kappa^{\prime} \frac{\theta (\kappa 
   -\kappa^{\prime})}{(\kappa -\kappa^{\prime})_+} \tilde{K}
 (\kappa^{\prime}, -{\bf b}_{\perp}). 
\end{eqnarray}
Therefore Eq.~(\ref{rgf}) is written as
\begin{equation}
\mu \frac{d}{d\mu} \tilde{f} (y,{\bf b}_{\perp},\kappa) = -\int dz
d\kappa^{\prime} \tilde{\gamma}_f (y,z,\kappa,\kappa^{\prime};
{\bf b}_{\perp}, {\bf p}_{\perp}) \tilde{f} (z,{\bf
  b}_{\perp},\kappa^{\prime}), 
\end{equation}
where
\begin{eqnarray} \label{tilrg}
\tilde{\gamma}_f (y,z,\kappa,\kappa^{\prime};
{\bf b}_{\perp}, {\bf p}_{\perp}) &=& -\frac{\alpha_s C_F}{\pi} \Bigl[
\Bigl(\frac{3}{2} \delta (y-z)+\frac{ -2 + (1+(y/z)^2) e^{i(1-y/z)
{\bf b}_{\perp} \cdot {\bf p}_{\perp}}}{(z-y)_+} \Bigr) \delta (\kappa
  -\kappa^{\prime})  \nonumber \\
&+& \frac{2\theta (\kappa -\kappa^{\prime})}{(\kappa
    -\kappa^{\prime})_+} \delta (y-z) \Bigr].
\end{eqnarray}
This is the main result for the renormalization group
behavior of the TMD collinear operator. For ${\bf b}_{\perp}=0$ in
Eq.~(\ref{tilrg}), we obtain the renormalization group equation for
the integrated PDF.

\section{Conclusion\label{conc}}
It is shown that the hadronic tensor, hence the scattering cross
section in  SIDIS is factorized into a hard part, the TMD PDF, and the
fragmentation function. The TMD PDF is further factorized into the TMD
collinear part and the TMD soft Wilson lines. The TMD PDF is
defined in terms of the matrix element of the gauge invariant TMD
collinear operator in which all the collinear particles are put on the
lightcone with no extra Wilson lines off the lightcone. The TMD soft
Wilson is defined in a similar way. The radiative correction for the
TMD PDF is computed with the zero-bin subtraction, and is shown
that the divergences of the form $1/(\eu\ei)$ are cancelled. This
cancellation is obtained by using the representation of a delta
function before the parameter $\epsilon$ in dimensional regularization
is set to zero. 
The resultant anomalous dimension kernel shows a
nontrivial dependence on the momentum fraction $x$ and the transverse
momentum ${\bf p}_{\perp}$ of the incoming parton. The radiative
correction for the 
TMD soft Wilson line is also performed and the anomalous dimension
kernel is the same as the integrated soft Wilson line except the delta
function for the transverse momentum. That is, the TMD soft Wilson
line does not scale with respect to the transverse momentum and it
shows the same scaling behavior as that of the integrated soft Wilson
line.

There are 
several issues in obtaining the TMD PDF and its renormalization group
behavior. In contrast to previous approaches, all the collinear
particles, hence all the collinear and soft Wilson lines are on the
lightcone. The cancellation of the divergence with the form $1/(\eu
\ei)$ in the radiative correction of the TMD quantities is explicitly
shown by carefully taking care of the limiting behavior of the
calculation in dimensional regularization. Since the transverse
momentum and
the $n$-component of the collinear momentum are closely related to
each other in dimensional regularization, if we put
the parameter $\epsilon$ in dimensional regularization to zero in the
first place, the calculation becomes inconsistent and the cancellation
does not occur, as claimed before. The technique used here is to
introduce external $p^2$ as an infrared cutoff and treat the
ultraviolet divergence in dimensional regularization. Though there is
an infrared cutoff, not all the infrared divergences are expressed in
terms of the cutoff, but there are some divergences which appear as
poles of $\ei$. It is important to disentangle all the types of these
divergences along with the ultraviolet divergence. After 
adding all the contributions, the infrared poles of $\ei$ are 
cancelled. Especially, the problematic divergence is of the form
$1/(\eu\ei)$ which should not be present in order for the theory to
make sense. In this approach there is no need
to introduce additional soft Wilson lines which offers a counterterm
to cancel the divergence from the virtual correction. Instead 
the use of the representation of a delta
function and the plus function prescription
are employed to show the cancellation of the unwanted
divergence. 

In evaluating the radiative corrections, the external nonzero $p^2$ is
introduced as an infrared cutoff and the limit $p^2 \rightarrow 0$ is
taken first before considering other limits. It is claimed that 
putting $p^2 =0$ before the calculation in dimensional regularization
would yield an incorrect result. Then a question can be raised
about whether the extraction of divergence is possible with the
on-shell scheme ($p^2 =0$) and treat both the infrared and the
ultraviolet divergences using dimensional regularization. This is in
principle possible, but it is extremely challenging to trace the
origin of the divergences. In this case, however, the calculation
with nonzero $p^2$ can be a guideline to trace the divergences. 

The zero-bin subtraction is also important in avoiding double
counting. In computing the radiative corrections of the collinear
operator, the integrals were computed naively including the unphysical
region with small label momentum. This region should be subtracted
because that region is covered by soft particles. Otherwise, double
counting occurs. The true soft contribution is decoupled from the
collinear part, and it appears in the soft Wilson line.

The renormalization group equation for the TMD collinear operator is a
complicated integro-differential equation in which the momentum
fraction $x$ appears in the delta function for the transverse
momentum. In impact parameter space, the renormalization group
equation involves an oscillating term which makes it difficult to
solve the equation. However the renormalization group equations
involve only physical variables. In contrast to previous
approaches, this equation does not involve the rapidity parameter
$\zeta$ \cite{Collins:1981uk}, or the mass parameters $m$ and
$\lambda$ \cite{Ji:2004wu}, therefore it is difficult to compare the
renormalization group equation or the radiative corrections directly.

The comparison between the TMD (or unintegrated) PDF or soft Wilson
lines and the integrated PDF or soft Wilson lines becomes manifest in
this approach. In this paper, it is shown by explicit 
calculation at one loop that the radiative corrections and the
renormalization group equations for the integrated PDF and soft Wilson
line are obtained by integrating those for the unintegrated (or TMD)
PDF and soft Wilson lines over the transverse momentum. This is
different from previous approach in which the comparison was not
transparent because the order of integrating over the transverse
momentum and the removal of the divergence makes the comparison
difficult \cite{Ji:2004wu}. However the calculational procedure in
this paper makes the comparison clear. 

Here  the TMD PDF is considered only for 
quark distribution functions. But it is straightforward to extend the
idea, and calculate the radiative corrections for gluon distribution
functions and their mixing at higher orders in $\alpha_s$. 
The analysis on the TMD collinear, soft operators
and their renormalization group behavior can be
extended to  other high-energy processes such as jet production or
Drell-Yan processes (for full QCD approach, see
Refs.~\cite{Collins:1981uk} and \cite{Collins:1982wa}). There will
appear appropriate collinear and soft 
operators, but the form and structure of these operators might be
different. It would be interesting to consider other high-energy
processes and to see if there are important processes in which the
transverse-momentum-dependent effects are relevant.

\section*{Acknowledgments}
The author is grateful to I.~W.~Stewart and S.~Fleming for discussion
and comments, and is
supported in part by the Korea Research Foundation Grant 
KRF-2005-015-C00103, and by funds provided by the U. S. Department of
Energy (D.O.E.) under cooperative research agreement
DE-FC02-94ER40818.

\end{document}